\pdfoutput=1
\documentclass[preprint,11pt,number,3p,times]{elsarticle}
\usepackage[latin1]{inputenc}

\bibstyle{elsarticle-num}

\usepackage[centertags]{amsmath}
\usepackage{amssymb}
\usepackage{epsfig}
\usepackage{varioref}
\usepackage{xspace}
\usepackage{bbm}
\usepackage{lscape}
\usepackage{paralist}
\usepackage{listings}
\usepackage{caption}
\DeclareCaptionFont{white}{\color{white}}
\DeclareCaptionFormat{listing}{%
  \parbox{\textwidth}{\colorbox{gray}{\parbox{\textwidth}{#1#2#3}}\vskip-4pt}}
\usepackage[T1]{fontenc}
\usepackage[scaled]{beramono}

\usepackage{textcomp}

\usepackage[usenames,dvipsnames,table]{xcolor}
\usepackage{fancyvrb}
\usepackage{microtype}

\newcommand\Small{\fontsize{9}{9.2}\selectfont}
\newcommand*\LSTfont{\Small\ttfamily\SetTracking{encoding=*}{-60}\lsstyle}

\newcommand{\code}[1]{{{\tt #1}}}
\newcommand{\codeq}[1]{\textquotedbl{\tt #1}\textquotedbl\xspace}
\newcommand{\pyrate}{PyR@TE\xspace}
\newcommand{\python}{\code{Python}\xspace}
\newcommand{\sarah}{\code{SARAH}\xspace}
\newcommand{\mathematica}{\code{Mathematica}\xspace}

\newcommand{\SO}[1]{\ensuremath{\mathrm{SO}(#1)}}
\newcommand{\SU}[1]{\ensuremath{\mathrm{SU}(#1)}}
\newcommand{\U}[1]{\ensuremath{\mathrm{U}(#1)}}

\newcommand{\bs}[1]{\ensuremath{\boldsymbol{#1}}}

\newcommand{\mtx}[4]{\ensuremath{\left(\begin{array}{cc}#1&#2\\#3&#4\end{array}\right)}}

\usepackage[colorlinks=true
,urlcolor=blue
,anchorcolor=blue
,citecolor=blue
,filecolor=blue
,linkcolor=black
,menucolor=blue
,linktocpage=true
]{hyperref}

\labelformat{section}{Section~#1}
\labelformat{subsection}{Section~#1}
\labelformat{equation}{Eq.~(#1)}
\labelformat{figure}{Fig.~#1}
\labelformat{subfigure}{Fig.~\thefigure#1}
\labelformat{table}{Tab.~#1}

\journal{Computer Physics Communications}

\AtBeginDocument{\labelformat{lstlisting}{Listing~#1}}

\begin{document}

\begin{frontmatter}

\begin{flushright}
\normalsize{LPSC-13-203}\\
\normalsize{BONN-TH-15}
\end{flushright}

\title{{\huge\bfseries PyR@TE}\\ Renormalization Group Equations for General Gauge Theories}

\author[lpsc]{F.~Lyonnet}
\author[lpsc]{I.~Schienbein}
\author[bonn]{F.~Staub}
\author[lpsc]{A.~Wingerter}

\address[lpsc]{Laboratoire de Physique Subatomique et de Cosmologie, UJF Grenoble 1, CNRS/IN2P3, INPG,\newline 53 Avenue des Martyrs, F-38026 Grenoble, France}
\address[bonn]{Bethe Center for Theoretical Physics and Physikalisches Institut der Universit\"at Bonn,
  Nussallee 12, 53115 Bonn, Germany}

\begin{abstract}
	Although the two-loop renormalization group equations for a general gauge field theory have been known for quite some time, deriving them for specific models has often been difficult in practice. This is mainly due to the fact that, albeit straightforward, the involved calculations are quite long, tedious and prone to error. The present work is an attempt to facilitate the practical use of the renormalization group equations in model building. To that end, we have developed two completely independent sets of programs written in \python and \mathematica{}, respectively. The \mathematica{} scripts will be part of an upcoming release of \sarah 4. The present article describes the collection of \python routines that we dubbed \pyrate which is an acronym for ``Python Renormalization group equations At Two-loop for Everyone''. In \pyrate, once the user specifies the gauge group and the particle content of the model, the routines automatically generate the full two-loop renormalization group equations for all (dimensionless and dimensionful) parameters. The results can optionally be exported to \LaTeX\ and \mathematica, or stored in a \python data structure for further processing by other programs. For ease of use, we have implemented an interactive mode for \pyrate in form of an \code{IPython Notebook}. As a first application, we have generated with \pyrate the renormalization group equations for several non-supersymmetric extensions of the Standard Model and found some discrepancies with the existing literature.
\end{abstract}

\end{frontmatter}

\section*{Program Summary}
\noindent{\em Program title:} \pyrate{}\\
{\em Program obtainable from:} {\tt http://pyrate.hepforge.org}\\
{\em Distribution format:}\/ tar.gz\\
{\em Programming language:} {\python}\\
{\em Computer:}\/ Personal computer\\
{\em Operating system:}\/ Tested on Fedora 15, MacOS 10 and 11, Ubuntu 12\\
{\em Dependencies:} \code{SymPy}, \code{PyYAML}, \code{NumPy}, \code{IPython}, \code{SciPy}\\
{\em Typical running time:}\/ Tens of seconds per model (one-loop), tens of minutes (two-loop)\\
{\em Nature of problem:}\/ Deriving the renormalization group equations for a general quantum field theory.\\
{\em Solution method:}\/ Group theory, tensor algebra

\clearpage
\newpage

\section{Introduction}
\label{sec:introduction}
The Standard Model (SM) is an impressively successful theory. 
It has been tested in a very large number of precision measurements
in low-energy experiments and at high-energy colliders,
and, despite all efforts, no solid evidence for physics beyond the Standard Model (BSM)
has emerged.
The recent discovery of a Higgs boson at the Large Hadron Collider at CERN \cite{Aad:2012tfa,Chatrchyan:2012ufa} is consistent with
this picture. Indeed, the couplings of this particle are in very good agreement with the
predictions from the SM, and its mass $m_H \simeq 126$ GeV also lies in the right ballpark anticipated by SM fits to electroweak precision data.

This particular value of the Higgs mass is quite intriguing
when analyzing the stability of the electroweak symmetry breaking vacuum and the
perturbativity of the underlying dynamics which involves running the couplings from the electroweak scale to higher energies
using the renormalization group equations (RGEs) \cite{Sher:1988mj}.
As a general rule, the $\beta$-function of the quartic Higgs coupling $\lambda$ receives positive (negative) contributions
from scalars (fermions), leading to an increasing (decreasing) contribution to the running of $\lambda$ with
increasing energy.
Clearly, we need $\lambda > 0$ to have a stable minimum in the Higgs potential.
However, slightly negative values of $\lambda$ are also admissible, if they lead to a metastable vacuum
with a lifetime which exceeds the age of the universe.
In the SM, given $m_H$ and the mass of the top quark (which gives the dominant fermionic contribution due to the large Higgs-top Yukawa coupling), one can ask at what scale $\lambda$ turns negative, thus implying an internal inconsistency and the breakdown of the (perturbative) SM.
A detailed analysis of this question depends on
\begin{inparaenum}[(i)]
\item the boundary conditions for the RGEs at the weak scale,
\item the running of the RGEs of the SM (at a given loop order), possibly modified by the presence of extra particles, and
\item the perturbative validity of the RGEs.
\end{inparaenum}
The parameters that have the largest effect on the boundary conditions are the Higgs and the top mass, and the strong coupling constant. Most interestingly, it turns out that with present data the values are just
right so that the SM with a (meta-)stable vacuum can be a consistent theory up to very high energies, potentially up to the Planck scale \cite{Holthausen:2011aa,Degrassi:2012ry}.

Of course, the internal consistency of a theory is just a necessary condition for its validity, 
and there are several reasons of different quality to go beyond the SM. Supersymmetry (SUSY) is one of the best-motivated extensions of the SM. At a technical level, it addresses the hierarchy problem by canceling the large corrections to the Higgs mass, and the seeming unification of gauge couplings may be indicative of a grand unified theory at a higher scale. However, the conspicuous absence so far of (low-energy) supersymmetry at the LHC\footnote{Clearly, it is much too early to discard the idea of TeV-scale supersymmetry, and the increase in center of mass energy from 8 TeV to 13 TeV will open up the possibility to discover some of the SUSY particles, if they are heavier than originally expected.} has rekindled the interest in non-supersymmetric extensions of the SM\footnote{For the by now standard motivation for SUSY, we refer the reader to the standard literature \cite{Nilles:1984ge,Haber:1984rc,Martin:1997ns}.}. 
The present work aims to provide a tool useful to explore BSM scenarios that go beyond supersymmetry.

In the context of SUSY several public codes exist which numerically evolve the RGEs not only 
for the minimal supersymmetric standard model, but also for the next-to-minimal supersymmetric 
standard model \cite{Ellwanger:2006rn}, for high-scale seesaw scenarios \cite{Porod:2003um,Porod:2011nf} 
or for $R$-parity violating models \cite{Allanach:2001kg,Allanach:2009bv}. 
In addition, the {\tt Mathematica} packages {\tt Susyno} \cite{Fonseca:2011sy} and
{\tt SARAH} \cite{Staub:2008uz,Staub:2009bi,Staub:2010jh,Staub:2012pb} 
allow since a few years an automated calculation of all $\beta$-functions for  SUSY models. 
In contrast, there has not been much effort so far to push also non-SUSY models to that level of automatization.

Here, we present a \python program that automatically generates the full two-loop renormalization group equations 
for all (dimensionless and dimensionful) parameters of a general gauge theory. The gauge group, the particle content and many other input parameters can be specified by the user by editing text files in an easy-to-understand format.  Once the RGEs for the theory at hand have been calculated by \pyrate, the results can optionally be exported to \LaTeX\ and \mathematica, or stored in a \python data structure for further processing by other programs. Also, for the convenience of the user, we have implemented an interactive mode in form of an \code{IPython Notebook}.

The general RGEs for non-supersymmetric gauge theories have been known at two-loop accuracy for about 30 years 
\cite{Machacek:1983tz,Machacek:1983fi,Machacek:1984zw,Jack:1982hf,Jack:1982sr,Jack:1984vj}. 
In developing \pyrate, all known typos in the original series of papers by Machacek and Vaughn have been taken into account\footnote{See Ref.~\cite{Luo:2002ey} and the appendix of Ref.~\cite{Wingerter:2011dk}.}, and the code has been validated against several known results in the literature (see \ref{sec:validation}). 
Also, independently of the \python program, \mathematica routines \cite{new_sarah} have been developed and cross-checked against the \pyrate, so that we feel confident to have eliminated most sources of possible errors that might affect the correctness of the RGEs.

The scope of \pyrate is not limited to exploring the stability of the electroweak vacuum, as we discussed in some detail in the first part of the introduction. 
First, extensions of the SM by weak scale dark matter have been studied in the literature \cite{Cheung:2012nb}.
Second, in cases where the scale $\Lambda$ is well below a possible unification scale or the Planck scale
the knowledge of the RGEs is necessary whenever the boundary conditions are defined at the unification
scale. In such a case it is also interesting to study the stability (and perturbativity) of the theory in
a similar way as it is done in the SM. Third, in split SUSY scenarios \cite{Giudice:2004tc} the low-energy spectrum is effectively non-supersymmetric, and therefore requires the most general RGEs.

The rest of this article is organized as follows. We explain in \ref{sec:download} the installation 
of \pyrate and describe in \ref{sec:Runningpyrate} how it can be used. In \ref{sec:innerworking}
we provide details about the calculations performed by \pyrate before we discuss the validation of the results in 
\ref{sec:validation}. We conclude in \ref{sec:conclusion}.
More information about the supported gauge groups, irreducible representations (irreps)
and already implemented models is given in the appendices.


\section{Download and Installation}
\label{sec:download}

\pyrate is free software under the copyleft of the \href{http://www.gnu.org/copyleft/gpl.html}{GNU General Public License} and can be downloaded from the following web page:
\begin{center}
\href{http://pyrate.hepforge.org}{http://pyrate.hepforge.org}
\end{center}

To install \pyrate, simply open a shell and type:
\begin{Verbatim}[numbers=left,xleftmargin=20pt,formatcom=\color{gray}]
cd $HOME
wget http://pyrate.hepforge.org/downloads/pyrate-1.0.0.tar.gz
tar xfvz pyrate-1.0.0.tar.gz
cd pyrate-1.0.0/
\end{Verbatim} 
\label{verb:gap_group_properties}

For definiteness, we will assume here and in the following that you want to install \pyrate in your home directory (cf.~line 1 in the listing above). Otherwise, simply replace \codeq{\$HOME} by a directory of your choice. At the time of writing the present article, \pyrate is available in the version 1.0.0, and later you may need to replace this by a more recent version number\footnote{All versions of \pyrate will be available in the ``Downloads'' section of our web page \cite{Pyrate13}.} (cf.~line 2). Unpacking the tar ball (line 3) will then create a subdirectory that contains \pyrate. We will describe how to run the program in \ref{sec:Runningpyrate}.

\pyrate has the following minimal software requirements:
\begin{itemize}
	\item \python $\geq$ 2.7.1\footnote{\pyrate was developed with \python 2.7.1 but should work with more recent versions with the exception of \python 3 for which it has not been tested.} \cite{Python}
	\item \code{NumPy} $\geq$ 1.7.1 \cite{dubois.hinsen.hugunin-1996-cp} and \code{SciPy} 0.12.0 \cite{scipy:2013} 
	\item \code{SymPy} $\geq$ 0.7.2 \cite{sympy:2013}
	\item \code{IPython} $\geq$ 0.12 \cite{PER-GRA:2007}
	\item \code{PyYAML} $\geq$ 3.10 \cite{PyYAML:2013}
\end{itemize}

Most of these packages ship with any standard Linux distribution and are by default pre-installed on your system, but in case they are not, you can easily install them. All but one are available in the standard repositories and can be installed by the respective package manager of your system, e.g.~\codeq{yum install SymPy} for a Fedora-based distribution and \codeq{apt-get install python-sympy} for a Debian-based one. For PyYAML, you have to visit its web page \cite{PyYAML:2013} and follow the installation instructions.

If \code{SymPy} 0.7.2 is not available for your system in the repositories (or not in the correct version\footnote{If \code{SymPy} 0.7.3 is available on your system, you can patch it so that it works with \pyrate. You can find detailed instructions on how to do this on our web page \cite{Pyrate13}.}), you can easily install it by downloading the source code from its web page \cite{sympy:2013}:
\begin{Verbatim}[numbers=left,xleftmargin=20pt,formatcom=\color{gray}]
wget https://github.com/sympy/sympy/archive/sympy-0.7.2.tar.gz
tar xfvz sympy-0.7.2
mv sympy-sympy-0.7.2/sympy $HOME/pyrate-1.0.0/
\end{Verbatim}

After unpacking the tarball (line 2), move the subdirectory \codeq{SymPy} to where \pyrate is installed (line 3). In the next section, we will explain in detail how to run \pyrate.

\section{Running PyR@TE}
\label{sec:Runningpyrate}

We will first describe how to run \pyrate from the command line and later explain in some detail the interactive mode in \ref{subsec:interactive}. Throughout this section, we will use the SM to illustrate how to use \pyrate, since it is the theory people are most familiar with. Also, for the SM the output of \pyrate can easily be compared to the literature.

\subsection{First Steps}
\label{subsec:commandline}

To run \pyrate, open a shell, change to the directory where it is installed and enter:

\begin{Verbatim}[xleftmargin=20pt,formatcom=\color{gray}]
python pyR@TE.py -m models/SM.model 
\end{Verbatim}

The option \codeq{-m} (or \codeq{-\xspace-Model}) is used to read in a model file, in this case the SM. For now, we defer the discussion of how to create a model file to \ref{subsec:modelfile} and proceed directly with the calculation of the RGEs. Because the calculations can be quite time-consuming, \pyrate does not calculate them by default. Rather, the user has complete freedom over the parts of the calculation he needs. For instance, to calculate the RGEs for the gauge, Yukawa or quartic couplings, one would add the options \codeq{-\xspace-Gauge-Couplings}, \codeq{-\xspace-Yukawas}, \codeq{-\xspace-Quartic-Couplings}, respectively, or alternatively \codeq{-gc}, \codeq{-yu} or \codeq{-qc}:

\begin{Verbatim}[xleftmargin=20pt,formatcom=\color{gray}] 
python pyR@TE.py -m ./models/SM.model -gc -yu -qc 
\end{Verbatim} 

After \pyrate terminates and the shell prompt reappears, the results of the calculation will be available in the newly created subdirectory \codeq{\$HOME/pyrate-1.0.0/results}. Specifically, the \LaTeX{} file \codeq{RGEsOutput.tex} contains the RGEs and a summary of the settings and of the model for which the calculation was done. We will discuss other forms of output later in \ref{subsec:output}. Before we go into those details, we would first like to give an exhaustive list of the options used to control \pyrate.

\subsection{Settings and options}
\label{subsec:settings}

\pyrate and the type of output it generates are controlled by various options that we have summarized in \ref{tab:options}. Alternatively, one can also obtain the complete list of options by typing
\begin{Verbatim}[xleftmargin=20pt,formatcom=\color{gray}] 
python pyR@TE.py --help
\end{Verbatim} 
at the shell prompt. Most options are self-explanatory, and we will therefore not go into any details at this point. In later sections, we will illustrate their use by providing examples.

As the number of options increases, it is more convenient to save all settings in a file which can then be passed to \pyrate instead of appending a long string of options\footnote{Note that we provide no default settings file and that you have to create your own one e.g.~by copying the lines given in \ref{lst:SM_settings}.}:

\begin{Verbatim}[xleftmargin=20pt,formatcom=\color{gray}]
python pyR@TE.py -f SMsets.settings
\end{Verbatim}

The input file \codeq{SMsets.settings} is written in YAML \cite{yaml:2013} which is a human-readable format for storing information that can also be easily accessed by a computer. The lines in this file have the following generic structure:

\begin{Verbatim}[xleftmargin=20pt,formatcom=\color{gray}]
keyword: value
\end{Verbatim}

Here, \codeq{keyword} is a keyword predefined in \pyrate, and \codeq{value} is either a path, a filename or a Boolean, i.e.~\codeq{True} or \codeq{False}. 
For example, a typical \codeq{SMsets.settings} file could look like this:
\begin{lstlisting}[basicstyle=\LSTfont, label=lst:SM_settings,caption=SMsets.settings]
# YAML 1.1
---
Model: ./models/SM.model
Gauge-Couplings: True
Quartic-Couplings: True
Yukawas: True
ScalarMass: False
Two-Loop: False
verbose: True
\end{lstlisting}

Note that 
\begin{inparaenum}[(i)]
\item strings need not be delimited by quotes,
\item you can only use spaces as whitespaces, i.e.~tabulators are not allowed, and
\item the space after ":" is mandatory. 
\end{inparaenum}
For the keys that can be used in the settings file, we refer the reader again to \ref{tab:options}.

\begin{table}[h!]
\caption{List of all options that can be used to control \pyrate.}
\label{tab:options}
\begin{center}
\renewcommand{\arraystretch}{1.1}
\scalebox{0.83}{
\begin{tabular}{|l|l|l|}
\hline
Option& Keyword \enspace | \enspace Default & Description\\\hline\hline
\code{-\xspace-Settings/-f}& -\enspace|\enspace - &Specify the name of a {\it .settings} file.\\\hline
\code{-\xspace-Model/-m}& \code{Model\enspace|\enspace -}&Specify the name of a {\it. model} file.\\\hline
\code{-\xspace-verbose/-v}&\code{verbose\enspace|\enspace False} &Set verbose mode.\\\hline
\code{-\xspace-VerboseLevel/-vL}&\code{VerboseLevel\enspace|\enspace Critical} &Set the verbose level: {\it Info, Debug, Critical}\\\hline
\code{-\xspace-Gauge-Couplings/-gc}&\code{Gauge-Couplings\enspace|\enspace False} &Calculate the gauge couplings RGEs. \\\hline
\code{-\xspace-Quartic-Couplings/-qc}&\code{Quartic-Couplings\enspace|\enspace False} &Calculate the quartic couplings RGEs.\\\hline
\code{-\xspace-Yukawas/-yu} &\code{Yukawas\enspace|\enspace False} &Calculate the Yukawa RGEs.\\\hline
\code{-\xspace-ScalarMass/-sm}& \code{ScalarMass\enspace|\enspace False}&Calculate the scalar mass RGEs.\\\hline
\code{-\xspace-FermionMass/-fm}&\code{FermionMass\enspace|\enspace False} &Calculate the fermion mass RGEs.\\\hline
\code{-\xspace-Trilinear/-tr}&\code{Trilinear\enspace|\enspace False}&Calculate the trilinear term RGEs.\\\hline
\code{-\xspace-All-Contributions/-a}&\code{all-Contributions\enspace|\enspace False} &Calculate all the RGEs.\\\hline
\code{-\xspace-Two-Loop/-tl}&\code{Two-Loop\enspace|\enspace False} &Calculate at two-loop order.\\\hline
\code{-\xspace-Weyl/-w}&\code{Weyl\enspace|\enspace True} &The particles are Weyl spinors.\\\hline
\code{-\xspace-LogFile/-lg}&\code{LogFile\enspace|\enspace True} &Produce a log file.\\\hline
\code{-\xspace-LogLevel/-lv}&\code{LogLevel\enspace|\enspace Info} &Set the log level: {\it Info, Debug, Critical} \\\hline
\code{-\xspace-LatexFile/-tex}& \code{LatexFile\enspace|\enspace RGEsOutput.tex}&Set the name of the \LaTeX{} output file.\\ \hline
\code{-\xspace-LatexOutput/-texOut}&\code{LatexOutput\enspace|\enspace True} &Produce a \LaTeX{} output file.\\\hline
\code{-\xspace-Results/-res}&\code{Results\enspace|\enspace ./results} &Set the directory of the results\\\hline
\code{-\xspace-Pickle/-pkl}&\code{Pickle\enspace|\enspace False} &Produce a pickle output file.\\ \hline
\code{-\xspace-PickleFile/-pf}&\code{PickleFile\enspace|\enspace RGEsOutput.pickle} &Set the name of the pickle output file.\\\hline
\code{-\xspace-TotxtMathematica/-tm}&\code{ToM\enspace|\enspace False} &Produce an output to Mathematica.\\\hline
\code{-\xspace-TotxtMathFile/-tmf}&\code{ToMF\enspace|\enspace RGEsOutput.txt} &Set the name of the Mathematica output file.\\\hline
\code{-\xspace-Export/-e}&\code{Export\enspace|\enspace False}&Produce the numerical output.\\\hline
\code{-\xspace-Export-File/-ef}&\code{ExportFile\enspace|\enspace BetaFunction.py}&File in which the beta functions are written.\\\hline
\end{tabular}
}
\end{center}
\end{table}

\subsection{Implementing your own model}
\label{subsec:modelfile}

The previous sections described how to run \pyrate to calculate the RGEs for a given model. In this section we will explain how to create your own model file that you can use with \pyrate. As before, we will use the SM as an epitome to explain the format of the model file. In \ref{app:sample_model_files}, we give several examples of model files for various extensions of the SM. These examples and many more are also available in the \codeq{models} subdirectory that ships with \pyrate. 

The three ingredients needed to define a model file are the gauge group, the particle content and the scalar potential. The general form of the model file is similar to that of the settings file already described in \ref{subsec:settings}. Consider the following model file given in \ref{lst:SM_model_1}. The first line indicates that this is a YAML file. Lines 3-5 indicate the author of the model, the filename and the date when it was created.

\begin{lstlisting}[basicstyle=\LSTfont, label=lst:SM_model_1,caption=models/SM.model]
# YAML 1.1
---
Author: Florian Lyonnet
Date: 11.04.2013
Name: SM
Groups: {'U1': U1, 'SU2L': SU2, 'SU3c': SU3}

##############################
#Fermions assumed weyl spinors
##############################
Fermions: {
   Qbar: {Gen: 3, Qnb:{ U1: -1/6, SU2L: -2, SU3c: -3}},
   Lbar: {Gen: 3, Qnb:{ U1: 1/2, SU2L: -2, SU3c: 1}},
   uR: {Gen: 3, Qnb:{ U1: 4/6, SU2L: 1, SU3c: [1,0]}},
   dR: {Gen: 3, Qnb:{ U1: -1/3, SU2L: 1, SU3c: 3}},
   eR: {Gen: 3, Qnb:{ U1: -1, SU2L: 1, SU3c: 1}}
}

#############
#Real Scalars
#############

RealScalars: {
  Pi: {U1: 1/2, SU2L: 2, SU3c: 1},
  Sigma: {U1: 1/2, SU2L: 2, SU3c: 1},
}

##########################################################################
#Complex Scalars : have to be expressed in terms of Real Scalars see above
##########################################################################

CplxScalars: {
  H: {RealFields: [Pi,I*Sigma], Norm: 1/Sqrt(2), Qnb : {U1: 1/2, SU2L: 2, SU3c: 1}},
  H*: {RealFields: [Pi,-I*Sigma], Norm: 1/Sqrt(2), Qnb : {U1: -1/2, SU2L: -2, SU3c: 1}}
}

Potential: {

#######################################
# All particles must be defined above !
#######################################
 
Yukawas:{
  'Y_{u}': {Fields: [Qbar,uR,H*], Norm: 1},
  'Y_{d}': {Fields: [Qbar,dR,H], Norm: 1},
  'Y_{e}': {Fields: [Lbar,eR,H], Norm: 1}
 },
QuarticTerms: {
 \Lambda_1 : {Fields : [H,H*,H,H*], Norm : 1/2}
 },
ScalarMasses: {
 \mu_1 : {Fields : [H*,H], Norm : 1}
 }
}
\end{lstlisting}

On line 6 you find the definition of the gauge group labelled by the keyword \codeq{Groups}. The gauge group is a product of simple Lie algebras and any number of \U{1} factors (note, however, that we have not implemented kinetic mixing between the \U{1} factors). In turn, each simple Lie algebra or \U{1} is associated with a user-defined name (e.g.~\codeq{SU3c} on line 6), and a predefined \pyrate keyword that specifies the Lie algebra as a mathematical object (cf.~$\SU{3}$ on line 6). So far, we have implemented $\SU{N}$ for $N=2,\ldots,6$ and \U{1}, and in \ref{app:irreps} we present a list of irreducible representations (irreps) that are currently recognized by \pyrate. Note that this list will be extended in future versions of \pyrate. 

Next, we discuss how to add particles to our model (lines 11-35 in \ref{lst:SM_model_1}). We distinguish between \codeq{Fermions}, \codeq{RealScalars} and \codeq{CplxScalars}. Each particle is defined by giving it a name and then listing all its quantum numbers, cf.~e.g.~line 12 in \ref{lst:SM_model_1}:
\begin{Verbatim}[xleftmargin=20pt,formatcom=\color{gray}]
Qbar: {Gen: 3, Qnb:{ U1: -1/6, SU2L: -2, SU3c: -3}}
\end{Verbatim}
Here, \codeq{Gen} is a predefined keyword denoting the number of generations, but the names for the gauge group factors correspond to those specified by the user on line 6. The number of generations for a given particle can in principle be kept general, but then \pyrate will not be able to perform some basic simplifications and the result may look more complicated. The gauge quantum numbers can either be specified by the dimension of the corresponding irrep\footnote{For simple gauge group factors we use a minus sign to distinguish between a representation and its complex conjugate one. For a \U{1} factor the quantum number corresponds to the usual \U{1} charge in some physics normalization.}, or their Dynkin labels (see definition of \codeq{uR} on line 14). This is possible for all simple gauge groups, but for $\SU{2}$ we have to use a slightly more complicated notation, since we need to distinguish between a given representation and its complex conjugate\footnote{In \SU{2} any representation is equivalent to its complex conjugate one, but for contracting the \SU{2} indices this change of basis matters.}: \textquotedbl$(n-1,)$\textquotedbl\xspace will correspond to the $n$-dimensional representation, and \textquotedbl$(n-1,\mathrm{True})$\textquotedbl\xspace to its complex conjugate. Note that internally all the quantum numbers are translated to Dynkin labels, so if the dimension of a given irrep does not define it uniquely, the user has to use the Dynkin labels. A table with all the irreps that can be used in \pyrate is given in \ref{app:irreps}.

Let us now turn to discussing how to add scalars (lines 23-35 in \ref{lst:SM_model_1}). Real scalars are declared by using the keyword \codeq{RealScalars} and then specifying their gauge quantum numbers (lines 24-25). For complex scalars one has to declare the real degrees of freedom following the keyword \codeq{RealScalars} as before, and also group together the related degrees of freedom using the keyword \codeq{CplxScalars} (see lines 33-34). The user can choose a convenient normalization for the complex scalar using the keyword \codeq{Norm}. Also note that you have to declare $H^*$ explicitly (see line 34).

We mention in passing that in order to simplify the notation we have introduced a short-hand syntax. The preceding declarations (lines 1-35 in \ref{lst:SM_model_1}) can also be rewritten in the form given in \ref{lst:SM_model_6}.

\begin{lstlisting}[basicstyle=\LSTfont, label=lst:SM_model_6,caption=Short-hand syntax for the SM model file]
Groups: [U1,SU2,SU3]
Fermions: {
   Qbar: [3, -1/3, -2,-3],
   Lbar: [3, 1,-2,1],
   uR: [3,4/3,1, [1,0]],
   dR: [3,-2/3, 1,3],
   eR: [3,-2,1,1]
}
RealScalars: {
   Pi: [1,2,1],
   Sigma: [1,2,1]
}
CplxScalars: {
  H: {RealFields: [Pi,I*Sigma], Norm: 1/Sqrt(2), Qnb : [1, 2,  1]},
  H*: {RealFields: [Pi,-I*Sigma], Norm: 1/Sqrt(2), Qnb : [-1,-2,1]}
}
\end{lstlisting}

We now come to the potential which is introduced by the keyword \codeq{Potential} (lines 37-54 in \ref{lst:SM_model_1}) and has five parts, each preceded by its own keyword: Yukawa interactions (\codeq{Yukawas}), quartic terms (\codeq{QuarticTerms}), scalar masses (\codeq{ScalarMasses}), trilinear interactions (\codeq{TrilinearTerms}), and fermion masses (\codeq{FermionMasses}). Each term in one of the five parts is represented by a coupling constant (e.g.~\textquotedbl\verb|mu_1|\textquotedbl\xspace on line 52), a number of fields (\codeq{[H*,H]}) and a numerical factor (\codeq{Norm : 1}). Note that for the coupling constant we can 
use \LaTeX\xspace notation\footnote{In this case, quotation marks must be used so that the string is recognized as a latex expression.}which will then be used for the output.

\subsection{Output}
\label{subsec:output}

In this section we explain in some more detail the various formats in which \pyrate can generate output.

\paragraph{\LaTeX}
With the option \codeq{-\xspace-Latex-Output}, \pyrate generates a \LaTeX{} file whose name can be set by \codeq{-\xspace-LatexFile} followed by a filename. This is the most convenient way to obtain the RGEs in a human-readable format. The file will be saved in the directory specified by the option \codeq{-\xspace-Results}, or, more conveniently, set in a settings file (cf.~\vref{lst:SM_settings}).

\paragraph{Pickle}
As the name suggests, Pickle is used to efficiently store Python data structures (in our case the partial or full results of our calculations) for later use. It is particularly useful when combined with the interactive mode to be described in \ref{subsec:interactive}. We refer the reader to \ref{tab:options} for a short description of the options \codeq{-\xspace-Pickle} and \codeq{-\xspace-PickleFile}.

\paragraph{Mathematica}
To export results to Mathematica, \pyrate can produce a text file with lines that can be directly copy-pasted into a Mathematica notebook. This option is controlled by the switches \codeq{-\xspace-TotxtMathematica} and  \codeq{-\xspace-TotxtMathFile} (see \ref{tab:options} for more details).

\paragraph{Numerical evaluation}
The RGEs generated by \pyrate can be directly solved and visualized in one of two ways: Either from within \python or in a Mathematica notebook. We will describe in turn both approaches.

The option \codeq{-\xspace-TotxtMathematica} also produces a file\footnote{If the filename is not set by \codeq{-\xspace-TotxtMathFile}, the default name \codeq{RGEsOutput.txt\_numerics.m} will be chosen.} that ends on \codeq{\_numerics.m} that contains the equations as well as the information required by \mathematica to solve the RGEs. The package \codeq{RunPyRate\_RGEs.m} which is included in the directory \codeq{Source/Output} prepares the equations for \mathematica and uses its internal routines to solve the system. For instance, one would enter the following lines in \mathematica:

\begin{Verbatim}[numbers=left,xleftmargin=20pt,formatcom=\color{gray}]
PATH = "$HOME/pyrate-1.0.0/";
Get[PATH <> "results/RGEsOutput.txt_numeric.m"];
Get[PATH <> "/Source/Output/RunPyRate_RGEs.m"];
IncludeOffDiagonal=True;
\end{Verbatim}

Line 1 tells \mathematica where \pyrate is installed. Line 2 points to the file where the results are stored, and line 3 loads the package to solve the RGEs. The switch \codeq{IncludeOffDiagonal=True} instructs \mathematica to include the full matrix structure of the parameters in solving the RGEs and not to neglect off-diagonal entries. By contrast, \codeq{IncludeOffDiagonal=False} will neglect the off-diagonal terms. After the initialization, a 
routine called \codeq{RunRGEs} is available that takes as input the starting and ending points of the interval over which the RGEs are to be integrated as well as the initial values of the parameters:
\begin{Verbatim}[xleftmargin=20pt,formatcom=\color{gray}]
running=RunRGEs[3, 16, {g1->0.36, gSU2L->0.65, gSU3c->1.08}];
\end{Verbatim}
The first and second inputs are the logarithms of the scales where the running starts and ends, respectively. If Landau poles appear, \mathematica will terminate before reaching the end point. The third input is the initialization of the parameters that have non-zero values at the starting scale. For instance, to run the gauge couplings in the SM from 1~TeV to $10^{16}$~GeV and to plot the result, simply enter:
\begin{Verbatim}[xleftmargin=20pt,formatcom=\color{gray}]
Plot[{g1[x],gSU2L[x],gSU3c[x]} /.running[[1]],{x,3,16}];
\end{Verbatim}
This example is also included in the file \codeq{Example.nb} inside the \pyrate directory.

Now we explain how to run the RGEs from within \python. With the options \codeq{-\xspace-Export} and \codeq{-\xspace-Export-File}\footnote{If this option is skipped, the file will be named \codeq{BetaFunction.py} by default.} \pyrate creates two files: The first one contains the results of the calculation in a form that is amenable to numerical analysis (i.e. \code{NumPy} objects). The second one, named \verb|SolveRGEs.py|, is a \python script that solves\footnote{We use python.scipy.integrate \cite{scipy:2013} to numerically solve ordinary differential equations.} the RGEs stored in the first file and contains instructions on how to plot the results with Matplotlib. Note that the user is responsible for setting the interval over which the RGEs will be integrated (start and end points) and also the initial values of the parameters. The file contains comments that will guide the eye of the user to perform the necessary modifications.

In the next section we will introduce a user interface \`a la \mathematica, called an \code{IPython Notebook}, in which we can perform all the tasks described so far in an interactive way.

\subsection{Interactive \pyrate}

\label{subsec:interactive}
A very convenient and user friendly way of using \pyrate is to combine our code with an \code{IPython Notebook} \cite{PER-GRA:2007}. The first thing to do is to start it by typing in the \pyrate directory\footnote{The \code{IPython Notebook} is included in recent installations of \code{ipython}. If you are using an older version, you can download it from its web page \cite{PER-GRA:2007} or use the command \codeq{pip install ipython} (recommended) which should also take care of possible dependencies. If not pre-installed, the package manager \codeq{pip} can be installed by hand or using \codeq{easy\_install pip}.}:
\begin{Verbatim}[numbers=left,xleftmargin=20pt,formatcom=\color{gray}]
cd $HOME/pyrate-1.0.0/
ipython notebook
\end{Verbatim}
The \code{IPython Notebook} will then start in your default browser, and you will see all the available notebooks that are located in the \pyrate directory. You can now start executing one of these notebooks by simply clicking on the link.

\subsection{Pitfalls}
\label{sec:pitfalls}

There are some subtleties in the implementation of a model to which we would like to draw the user's attention. We start with commenting on the restrictions concerning the input format. To  ascertain that \python interprets all parts of the model file correctly, the user must make sure that:
\begin{enumerate}[(a)]
 \item only spaces and line breaks can be used as whitespaces, i.e.~tabulators are not allowed,
 \item there is a space after each colon,
 \item each element in any input file except for the last one should be separated by a comma. 
\end{enumerate}
In addition, beware that no operation on the fields is recognized, i.e.~for complex conjugated fields one needs to introduce a new symbol. For complex scalars, the real degrees of freedom have to be defined together with the required normalization. The Yukawa matrices are assumed to be symmetric in generation space. Therefore, if e.g.~some Yukawa terms are antisymmetric, \pyrate will return zero.

Finally, all indices are contracted automatically by \pyrate. For this purpose a database with the most common Clebsch-Gordon coefficients (CGCs) has been created, see \ref{app:irreps}. This database uses the following conventions:

\paragraph{Normalization} We assume a set of $n$ fields  $\phi_i$ with dimensions $D_i$ under an $\SU{N}$ gauge group. We will denote the CGC that gives the contraction of indices to an invariant combination as ${\cal C}$, i.e.
 \begin{equation}
	 {\cal C}_{i_1 i_2 \dots i_n}   \phi_{i_1} \phi_{i_2} \dots \phi_{i_n}.
 \end{equation}
does not transform under \SU{N}. Here, the $i_x$, $x=1,\ldots,n$ are the charge indices with respect to the gauge group. In contrast to {\tt Susyno} which has been used to create the database of CGCs we use a different normalization. Our convention is that 
\begin{equation}
	\sum_{i_1=1}^{D_1} \sum_{i_2=1}^{D_2} \dots \sum_{i_n=1}^{D_n} |{\cal C}_{i_1 i_2 \dots i_n}|^2 = \max(D_i).
\end{equation}
With this normalization we reproduce for instance the standard CGCs for all bilinear terms, but also those for $\SU{2}_L$ triplets with non-zero hypercharge and color sextets. However, we do not distinguish between $\SU{2}_L$ triplets with and without hypercharge and use the same CGCs for both of them. Therefore, our convention for triplets without hypercharge is different to the standard one by a factor $1/\sqrt{2}$. 
\paragraph{Conjugate irreps} In general, conjugate irreps are either defined by the corresponding Dynkin indices or by a negative dimension. However, we would like to stress that
\begin{enumerate}[(a)]
\item the ${\bf 2}$ under \SU{2} is related to its conjugate representation \bs{2^*}. Nevertheless, it is possible to use \codeq{-2} to represent \bs{2^*} which is then treated as a doublet with an additional $i \sigma_2$. For instance, the tensor product $\bs{2^*}\otimes \bs{2}$ is contracted with the Kronecker $\delta_{ij}$ whereas $\bs{2}\otimes \bs{2}$ by the anti-symmetric tensor $\epsilon_{ij}$. This shows up e.g.~in the case of the SM Yukawa couplings $Y_d$ and $Y_u$.
\item For self-conjugate representations like the adjoint ones, there are two ways to obtain a gauge singlet. To distinguish these two cases it is possible to use {\tt -A} as dimension of the adjoint of $\SU{N}$. The convention is then that bilinear terms of the form $\bs{A}^*\otimes \bs{A}$ are always contracted with a Kronecker $\delta_{ij}$, while for $\bs{A}\otimes \bs{A}$ the CGCs as calculated by {\tt Susyno} are used. For instance, $\bs{3}\otimes \bs{3}$ in \SU{2} is contracted by a matrix of the form 
\begin{equation}
 \left(\begin{array}{ccc} 0 & 0 & 1 \\ 0 & -1 & 0 \\ 1 & 0 & 0 \end{array}\right)
\end{equation}
while for $\bs{3}^*\otimes \bs{3}$ the three-dimensional identity matrix is used.
\end{enumerate}

\section{Calculating the RGEs: A Summary}
\label{sec:innerworking}
We are going to present in this section details of the 
calculation performed by \pyrate. As first step we show 
how the fundamental information, the generators of the 
gauge groups and the CGCs are derived. Afterwards we re-write 
the results presented in  
Refs.~\cite{Machacek:1983tz,Machacek:1984zw} 
in a more explicit form which is more suitable for 
building up algorithms. \\
\subsection{Clebsch Gordan coefficient and Generators}
In order to perform the calculation the group properties of each representation have to be known. This includes the value of the quadratic Casimir operator, the Dynkin index as well as an explicit matrix representation of the irreducible representations. Moreover, in order to build a gauge invariant potential, the relevant CGCs have to be used for each one of the terms. In order to do so we developed a database of all the bilinear, trilinear as well as quadratic invariants that can appear in the contraction of irreps allowed in \pyrate\footnote{
For a complete list of the supported irreps see \ref{app:irreps}.}.
This database was constructed using the \mathematica package \code{Susyno 2.0} \cite{Fonseca:2011sy} from which we also extracted the matrix representation of the quadratic Casimir as well as the Dynkin index for all the irreps.

\subsection{Generators}
\label{sec:generators}
The covariant derivatives for the fermion $\Psi$ and real scalar $\Phi$\footnote{If not stated otherwise,
we are going to assume that fermions are always described by Weyl spinors and scalars are real.} 
 are written as
\begin{eqnarray}
D_\mu \Psi_a &=& \partial_\mu \Psi_a - i g t^A_{ab} \Psi_b 
\,,\\
D_\mu \Phi_a &=& \partial_\mu \Phi_a - i g \Theta^A_{ab} \Phi_b 
\, .
\end{eqnarray}
Here, the index $A$ runs from 1 to the dimension of the Lie group, the indices $a$, $b$ from 1 to the dimension of the irrep
under which the field transforms and $\mu$ is a Lorentz index. 

\subsubsection*{Generators for scalars}
In the calculation the scalar fields are assumed to be real and the $\Theta^A$ matrices to be purely imaginary and anti-symmetric. To get the correct form of these generators we can start with a complex scalar $\varphi$ which transforms similar to the complex fermion as
\begin{equation}
\varphi \to e^{i \epsilon^A t^A} \varphi \, .
\end{equation}
We can now define a real vector $\Phi$ which consists of the real and imaginary component of $\varphi$
\begin{equation}
\Phi = \left(\begin{array}{c}\mbox{Re}(\varphi) \\ \mbox{Im}(\varphi) \end{array}\right) \, . 
\end{equation}
$\Phi$ transforms according to 
\begin{equation}
\varPhi \to e^{i \epsilon^A \Theta^A} \varPhi \,,
\end{equation}
from where we can obtain the relation 
\begin{equation}
\label{eq:ScalarGenerators}
\Theta^A = i \left(\begin{array}{cc} \text{Im}(t^A) & \text{Re}(t^A) \\ -\text{Re}(t^A) & \text{Im}(t^A) \end{array}   \right)  \, .
\end{equation}
We can demonstrate this construction at the example of the fundamental representation of $\SU{2}$. Note, that this is equivalent to embedding 
$\SU{2}$ into $\SO{4}$. The complex, Hermitian generators for $\SU{2}$ are proportional to the Paul matrices
\begin{equation}
\{\sigma_{1},\sigma_{2},\sigma_{3}\}=\left\{\mtx{0}{1}{1}{0},\mtx{0}{-i}{i}{0},\mtx{1}{0}{0}{-1}\right\}\, .
\end{equation}
Now, applying the relation \ref{eq:ScalarGenerators} to all three matrices, we obtain the following set of generators
\begin{eqnarray}
	\{\tilde{\Sigma}_{1},\tilde{\Sigma}_{2},\tilde{\Sigma}_{3}\} = i \enspace \left\{\left(\begin{array}{cccc}0&0&0&1\\0&0&1&0\\0&-1&0&0\\-1&0&0&0\end{array}\right),\left(\begin{array}{cccc}0&-1&0&0\\1&0&0&0\\0&0&0&-1\\0&0&1&0\end{array}\right), \left(\begin{array}{cccc}0&0&1&0\\0&0&0&-1\\-1&0&0&0\\0&1&0&0\end{array}\right)\right\}\, .
\end{eqnarray}
These matrices are indeed antisymmetric and imaginary and we can check that they satisfy the same
commutation relations as the $\sigma$'s:
\begin{eqnarray}
	\left[ \tilde{\Sigma}_{1},\tilde{\Sigma}_{2} \right] &=&  2 i \tilde{\Sigma}_{3},	\\
	\left[ \tilde{\Sigma}_{1},\tilde{\Sigma}_{3} \right] &=& - 2 i \tilde{\Sigma}_{2},	\\
	\left[ \tilde{\Sigma}_{2},\tilde{\Sigma}_{3} \right] &=&  2 i \tilde{\Sigma}_{1}\, .
\end{eqnarray}

\subsection{Gauge interactions}
\label{sec:how_gauge}

After we have prepared all information we need, we can start with the discussion 
how to calculate the $\beta$-functions. As a first step we concentrate
on the terms involving only gauge interactions. The 
basic objects to calculate the one- and two-loop $\beta$-functions for the gauge 
couplings in absence of any matter interaction are the quadratic Casimir operator $C_2$ and the Dynkin 
index $S_2$ of the gauge group. Those indices can be related to the generators $t^A$ for fermions 
and $\Theta^A$ for scalars introduced in \ref{sec:generators}
\begin{eqnarray}
&C_2^{ab}(S) = \Theta^A_{ac} \Theta^A_{cb}\,,\hspace{1cm} S_2(S)\delta_{AB}= \mbox{Tr}(\Theta^A\Theta^B) \, , &\\
&C_2^{ab}(F) = t^A_{ac} t^A_{cb}\,,\hspace{1cm} S_2(F)\delta_{AB}= \mbox{Tr}(t^A t^B) \, . &
\end{eqnarray}
The first step is to make the meaning of the indices more explicit. For this purpose we 
assume that we have a gauge sector which is a direct product of $n$ non-Abelian gauge groups
and at most one Abelian gauge group $\U{1}$. The non-Abelian groups are labeled with small letters:
$\U{1} \times {\cal G}_1 \times \dots \times {\cal G}_k \times \dots \times {\cal G}_n$. 
In the case of several $\U{1}'s$ the situation is 
more involved due to the impact of kinetic mixing \cite{Holdom:1985ag}. Rules to derive the entire 
two-loop RGEs in this context have just recently been given in Ref.~\cite{Fonseca:2013bua}.

For the charge indices with respect to the non-Abelian gauge groups we are going to use 
Greek letters in the following. In addition, there are sets of fermion fields $\psi^1$ \dots $\psi^{n_f}$ 
and real scalars $\phi^1$ \dots $\phi^{n_s}$ which can be charged under these gauge groups. 
Moreover, all fields can come in an arbitrary number of generations ${\cal N}_F^i$ respectively ${\cal N}_S^i$ 
so that in general each field carries $n+1$ indices. 
Using these conventions, we can rewrite the group constants
for one particular non-Abelian gauge group $k$ as 
\begin{eqnarray}
C_{2,k}^{\psi^i_{g_i,\alpha_1 \dots \alpha_k \dots \alpha_n} \psi^j_{g_j,\beta_1 \dots \beta_k \dots \beta_n}}(F) &=& \delta_{ij} \delta_{g_i g_j} \delta_{\alpha_1 \beta_1} \dots \delta_{\alpha_{k-1} \beta_{k-1}} \delta_{\alpha_{k+1} \beta_{k+1}} \dots \delta_{\alpha_n \beta_n} {\cal C}_{k}(\Lambda(\psi^i)) \, , \\
S_{2,k}^{\psi^i_{g_i,\alpha_1 \dots \alpha_k \dots \alpha_n} \psi^j_{g_j,\beta_1 \dots \beta_k \dots \beta_n}}(F) &=& \delta_{ij} \delta_{g_i g_j} \delta_{\alpha_1 \beta_1} \dots \delta_{\alpha_{k-1} \beta_{k-1}} \delta_{\alpha_{k+1} \beta_{k+1}} \dots \delta_{\alpha_n \beta_n} {\cal S}_{k}(\Lambda(\psi^i)) \,,
\end{eqnarray}
and similar for scalars. Here, we introduced ${\cal C}_k(\Lambda)$ and ${\cal S}_k(\Lambda)$ which are the quadratic Casimir and Dynkin index 
of an irrep with highest weight $\Lambda$ with respect to the gauge group $k$. ${\cal C}$ can be calculated 
using the well-known formula 
\begin{equation}
{\cal C}(\Lambda) =  \langle\Lambda, \Lambda + 2 \delta \rangle \, ,
\end{equation}
with $\delta=(1,1,\dots,1)$ in the Dynkin basis. The Dynkin index is normalized in a way that the value for the fundamental irrep is $\frac{1}{2}$:
\begin{equation}
{\cal S}_k(\Lambda) = \frac{N_k(\Lambda)}{N({\cal G}_k)} {\cal C}_k(\Lambda) \, .
\end{equation}
Here, $N(\Lambda)$ is the dimension of the irrep and $N({\cal G})$ the dimension of the adjoint representation. 
For the Abelian gauge group we have 
\begin{eqnarray}
C_{2,U(1)}^{\psi^i_{g_i,\alpha_1 \dots \alpha_k \dots \alpha_n} \psi^j_{g_j,\beta_1 \dots \beta_k \dots \beta_n}}(F) = S_{2,U(1)}^{\psi^i_{g_i,\alpha_1 \dots \alpha_k \dots \alpha_n} \psi^j_{g_j,\beta_1 \dots \beta_k \dots \beta_n}}(F)= \delta_{ij} \delta_{g_i g_j} \delta_{\alpha_1 \beta_1} \dots  \delta_{\alpha_n \beta_n} Q(\Psi^i)^2 \,  .
\end{eqnarray}
$Q$ is the charge of the field which might include a GUT normalization.  
We can now define the Dynkin index summed over all states present in the model:
\begin{eqnarray}
\tilde{S}_{2,k}(S) &=& \sum_{s=1}^{n_s} \prod_{l=1}^n {\cal N}^s_S \tilde{N}(\Lambda(s))_{lk} {\cal S}_{k}(\Lambda(s)) \, ,\\
\tilde{S}_{2,k}(F) &=& \sum_{f=1}^{n_f} \prod_{l=1}^n {\cal N}^f_F \tilde{N}(\Lambda(f))_{lk} {\cal S}_{k}(\Lambda(f)) \, ,
\end{eqnarray}
with 
\begin{equation}
 \tilde{N}(\Lambda)_{lk} = 
 \begin{cases}
 N_l(\Lambda) & \quad \text{if $l \ne k$},
 \\
 1 & \quad \text{else if $l=k$}\, .
 \end{cases}
 \end{equation}
For the Abelian gauge group we get 
\begin{eqnarray}
\tilde{S}_{2,U(1)}(S) &=& \sum_{s=1}^{n_s} \prod_{l=1}^n {\cal N}^s_S N_l(\Lambda(s)) Q(s)^2 \, ,\\
\tilde{S}_{2,U(1)}(F) &=& \sum_{f=1}^{n_f} \prod_{l=1}^n {\cal N}^f_F N_l(\Lambda(f)) Q(f)^2 \, .
\end{eqnarray}
With these results, the one-loop $\beta$ functions of a particular gauge coupling $g_{{\cal G}_k}$ is calculated via
\begin{equation}
\label{eq:betaG1}
\beta(g_{{\cal G}_k}) = -\frac{g_{{\cal G}_k}^3}{16\pi^2}\left(\frac{11}{3}C({\cal G}_k) - \frac{2}{3} \tilde{S}_{2,k}(F) - \frac{1}{6} \tilde{S}_{2,k}(S) \right)\, ,
\end{equation}
where $C(\cal G)$ is the quadratic Casimir operator in the adjoint representation.

We want to clarify these expression with the example of the SM, but concentrate for brevity just on the non-Abelian sector. That means, we have the gauge groups $\SU{2}_L \times \SU{3}_C$, the fermionic fields\footnote{In brackets we show the quantum numbers with respect to $\SU{2}_L \times \SU{3}_C$.}  
$q (2,3)$, $\bar{u}(1,\bar{3})$, $\bar{d}(1,\bar{3})$, $l(2,1)$, $e (1,1)$ and two real scalars $\phi_h (2,1)$, $\sigma_h (2,1)$ which are stemming from one 
complex Higgs doublet 
\begin{equation}
\label{eq:decompositioHiggs}
H = \frac{1}{\sqrt{2}} (\phi_h + i \sigma_h) \, .
\end{equation}
All fermions appear in ${\cal N}_G$ generations while we restrict the generation of Higgs fields to one. Hence, we obtain:
\begin{eqnarray}
\tilde{S}_{2,1}(F) =  {\cal N}_G \left[3 \cdot {\cal S}_{\SU{2}}(\Lambda(q)) + {\cal S}_{\SU{2}}(\Lambda(l)) \right] = 2 {\cal N}_G \,, \\
\tilde{S}_{2,1}(S) =  {\cal S}_{\SU{2}}(\Lambda(\phi^h)) + {\cal S}_{\SU{2}}(\Lambda(\sigma^h))  = 1 \, ,\\
\tilde{S}_{2,2}(F) =  {\cal N}_G \left[2 \cdot {\cal S}_{\SU{3}}(\Lambda(q))  + {\cal S}_{\SU{3}}(\Lambda(d)) + {\cal S}_{\SU{3}}(\Lambda(u))\right] = 2 {\cal N}_G \, .
\end{eqnarray}
In addition, ${\cal C}(\SU{N}) = N$ holds. Hence, we obtain from \ref{eq:betaG1}
\begin{eqnarray}
\beta(g_2) &=& - \frac{g_2^3}{16\pi^2}  \left(\frac{11}{3} 2 - \frac{2}{3} 2 {\cal N}_G - \frac{1}{6} \right) = - \frac{g_2^3}{16\pi^2} \left( \frac{43}{6} - \frac{4}{3} {\cal N}_G \right) \, ,\\
\beta(g_3) &=& - \frac{g_3^3}{16\pi^2}  \left(\frac{11}{3} 3 - \frac{2}{3} 2 {\cal N}_G\right) = - \frac{g_3^3}{16\pi^2}  \left(11 - \frac{4}{3} {\cal N}_G\right) \, .
\end{eqnarray}
Here, we have introduced the short form $g_2 = g_{\SU{2}}$ and $g_3 = g_{\SU{3}}$. We continue with the two-loop $\beta$-functions. We have to clarify the meaning of
\begin{equation}
|C({\cal G})|^2 \,,\hspace{1cm} S(R) C({\cal G}) \,,\hspace{1cm} S(R)C(R) \,,
\end{equation}
with $R=S,F$. The easy part is $|C({\cal G})|^2$ which results for a $\SU{N}$ gauge group in $N^2$. We can use 
the already introduced $\tilde{S}$ to express $S(R) C({\cal G})$ as
\begin{equation}
S(R) C({\cal G}) \to \tilde{S}_{2,k} C({\cal G}_k) \, . 
\end{equation}
Furthermore, the correct multiplicity for the term $S(R) C(R)$ can be obtained by inspecting a representative Feynman diagram. The result is
\begin{equation}
(S(R) C(R))_k \equiv \sum_{r} \sum_l  g^2_k g^2_l {\cal N}_r {\cal S}_{k}(\Lambda(r)) {\cal C}_{l}(\Lambda(r))\prod_{m}\tilde{N}(\Lambda(r))_{mk}
\end{equation}
with $r=s\ \text{if}\ R=S$ or $f\ \text{if}\ R=F$. Note, there is no (implicit) sum over $k$. 
Hence, the two-loop contributions stemming purely from gauge interactions to the $\beta$ functions are in general given by
\begin{eqnarray}
	\beta^{II}(g_{{\cal G}_k}) &=& - \frac{g_{{\cal G}_k}}{(16\pi^2)^2} \Bigg[ g_{{\cal G}_k}^4\frac{34}{3} |C({\cal G}_k)|^2 - \frac{1}{2} \left(4 (S(F) C(F))_{k} + \frac{20}{3} \tilde{S}_{2,k}(F) C({\cal G}_k)\right) \nonumber \\   
&& \hspace{2cm} - \left(2 (S(S) C(S))_k + \frac{1}{3} \tilde{S}_{2,k}(S) C({\cal G}_k) \right) \Bigg] \, .
\end{eqnarray}
For a $\SU{N}$ gauge group this can be simplified by using $|C({\cal G}_k)| = N$ and $|C({\cal G}_k)|^2 = N^2$.
For the same particle content as above we obtain
\begin{eqnarray}
(S(F) C(F))_1 &=& {\cal N}_G \left(\frac{3}{2} g_2^4 + 2 g_2^2 g_3^2 \right) \,,\\
(S(S) C(S))_1 &=& \frac{3}{4} g_2^4 \,,\\
(S(F) C(F))_2 &=& {\cal N}_G \left(\frac{3}{4} g_2^2 g_3^2 + \frac{8}{3} g_3^4 \right) \,,
\end{eqnarray}
and end up with the $\beta$ functions
\begin{eqnarray}
\beta^{II} (g_2) &=& -\frac{g_2}{(16 \pi^2)^2} \left[g_2^4 \frac{34}{3} 2^2 - \frac{1}{2}\left(4 {\cal N}_G\left(\frac{3}{2} g_2^4 + 2 g_2^2 g_3^2\right) + \frac{20}{3} \cdot 2 \cdot 2 \cdot {\cal N}_G \cdot g_2^4 \right) - \left(2 \cdot \frac{3}{4} g_2^4 + \frac{1}{3} \cdot 2 \cdot g_2^4\right)   \right] \nonumber \\ 
&& \hspace{1cm} = -\frac{g_2}{(16 \pi^2)^2} \left[\frac{138}{4} g_2^4 - {\cal N}_G \left(\frac{49}{3} g_2^4 + 4 g_2^2 g_3^2\right) - \frac{13}{6} g_2^4  \right] \,,\\
\beta^{II} (g_3) &=& -\frac{g_3}{(16 \pi^2)^2} \left[g_3^4 \cdot \frac{34}{3} \cdot 3^2 - \frac{1}{2}\left(4 {\cal N}_G \left(\frac{3}{4} g_2^2 g_3^2 + \frac{8}{3} g_3^4 \right) + \frac{20}{3} \cdot 2 \cdot 3 {\cal N}_G \cdot g_3^4 \right)   \right] \nonumber \\ 
&& \hspace{1cm} = -\frac{g_3}{(16 \pi^2)^2} \left[102 g_3^4 - {\cal N}_G \left(\frac{76}{3} g_3^4 + \frac{3}{2} g_2^2 g_3^2\right)  \right] \, .
\end{eqnarray}

\subsection{Matter interactions}
\label{sec:how_matter}
The potential 
of a general, renormalizable quantum field theory using the 
standard notation in the literature consists of the 
following terms:
\begin{eqnarray}
-V &=& \frac{1}{2} (Y^a_{jk}\Psi_j \xi\Psi_k \Phi_a - (m_f)_{jk} \Psi_j \xi \Psi_k + h.c.) \nonumber \\
 && +  \frac{1}{4!} \lambda_{abcd} \Phi_a\Phi_b\Phi_c\Phi_d- 
 \frac{1}{2} m_{ab}^2 \Phi_a \Phi_b - \frac{1}{3!} h_{abc} \Phi_a \Phi_b \Phi_c
 \label{eq:potential} \, ,
\end{eqnarray}
with $\xi = \pm i \sigma_2$. Note, a tadpole term $t \Phi$, which is in principle 
possible for a gauge singlet is not present, since it can always be absorbed into 
a shift of $\Phi$. Using the same conventions as introduced in \ref{sec:how_gauge}, 
we can re-write \ref{eq:potential} as
\begin{eqnarray}
-V &=& \frac{1}{2} \sum_{i,j,k} [C_{\alpha_1 \beta_1 \gamma_1} \cdot \dots  \cdot C_{\alpha_n \beta_n \gamma_n}] Y^{\phi^k_{g_1,\alpha_1 \dots \alpha_n}}_{\psi^i_{g_2, \beta_1 \dots \beta_n} \psi^j_{g_3, \gamma_1 \dots \gamma_n}} 
\phi^k_{g_1,\alpha_1 \dots \alpha_n} \psi^i_{g_2, \beta_1 \dots \beta_n} \xi \psi^j_{g_3, \gamma_1 \dots \gamma_n} + h.c.  \nonumber \\
&& -\frac{1}{2} \sum_{i,j} [C_{\alpha_1 \beta_1} \cdot \dots  \cdot C_{\alpha_n \beta_n}]  (m_f)_{\psi^i_{g_1, \alpha_1 \dots \alpha_n} \psi^j_{g_3, \beta_1 \dots \beta_n}} \psi^i_{g_1, \alpha_1 \dots \alpha_n} \xi \psi^j_{g_3, \beta_1 \dots \beta_n} + h.c.  \nonumber \\
&& +\frac{1}{4!} \sum_{i,j,k,l} [C_{\alpha_1 \beta_1 \gamma_1 \delta_1} \cdot \dots \cdot C_{\alpha_n \beta_n \gamma_n \delta_n}] \lambda_{\phi^i_{g_1,\alpha_1 \dots \alpha_n} \phi^j_{g_2,\beta_1 \dots \beta_n} \phi^k_{g_3,\gamma_1 \dots \gamma_n} \phi^l_{g_4,\delta_1 \dots \delta_n}} \phi^i_{g_1,\alpha_1 \dots \alpha_n} \phi^j_{g_2,\beta_1 \dots \beta_n} \phi^k_{g_3,\gamma_1 \dots \gamma_n} \phi^l_{g_4,\delta_1 \dots \delta_n} \nonumber \\
&& -\frac{1}{3!} \sum_{i,j,k} [C_{\alpha_1 \beta_1 \gamma_1} \cdot \dots \cdot C_{\alpha_n \beta_n \gamma_n}] h_{\phi^i_{g_1,\alpha_1 \dots \alpha_n} \phi^j_{g_2,\beta_1 \dots \beta_n} \phi^k_{g_3,\gamma_1 \dots \gamma_n} } \phi^i_{g_1,\alpha_1 \dots \alpha_n} \phi^j_{g_2,\beta_1 \dots \beta_n} \phi^k_{g_3,\gamma_1 \dots \gamma_n}  \nonumber \\
&&- \frac{1}{2} \sum_{i,j} [C_{\alpha_1 \beta_1} \cdot \dots \cdot C_{\alpha_n \beta_n}] m^2_{\phi^i_{g_1,\alpha_1 \dots \alpha_n} \phi^j_{g_2,\beta_1 \dots \beta_n} } \phi^i_{g_1,\alpha_1 \dots \alpha_n} \phi^j_{g_2,\beta_1 \dots \beta_n}  \, .
\label{eq:GeneralPotential}
\end{eqnarray}
Here, we introduced the CGC $C$ which vanish for combinations of fields which are not gauge invariant. In addition, to simplify the notation we kept also a charge ('dummy') index for fields which are not charged under a particular gauge group. In this case relations like $C_{\alpha \beta \gamma} = C_{\alpha \gamma}$ hold for a dummy index $\beta$, of course. Finally, we are going to define the following objects:
\begin{eqnarray}
\label{eq:defY}
{\cal Y}^{\phi^i_{g_i, \alpha_1 \dots \alpha_n}}_{\psi^j_{g_j, \beta_1 \dots \beta_n} \psi^k_{g_k, \gamma_1 \dots \gamma_n}} &=& -\frac{\partial^3 V}{(\partial \phi^i_{g_i, \alpha_1 \dots \alpha_n}) (\partial \psi^j_{g_j, \beta_1 \dots \beta_n}) (\partial \psi^k_{g_k, \gamma_1 \dots \gamma_n})} \,, \\
{\cal M}_{\psi^j_{g_j, \beta_1 \dots \beta_n} \psi^k_{g_k, \gamma_1 \dots \gamma_n}} &=& -\frac{\partial^2 V}{(\partial \psi^j_{g_j, \beta_1 \dots \beta_n}) (\partial \psi^k_{g_k, \gamma_1 \dots \gamma_n})} \,,\\
{\cal L}^{\phi^i_{g_i, \alpha_1 \dots \alpha_n} \phi^j_{g_j, \beta_1 \dots \beta_n} \phi^k_{g_k, \gamma_1 \dots \gamma_n} \phi^l_{g_l, \delta_1 \dots \delta_n}   } &=& -\frac{\partial^4 V}{(\partial \phi^i_{g_i, \alpha_1 \dots \alpha_n}) (\partial \phi^j_{g_j, \beta_1 \dots \beta_n} ) (\partial \phi^k_{g_k, \gamma_1 \dots \gamma_n}) (\partial \phi^l_{g_l, \delta_1 \dots \delta_n})} \,,\\
{\cal H}^{\phi^i_{g_i, \alpha_1 \dots \alpha_n} \phi^j_{g_j, \beta_1 \dots \beta_n} \phi^k_{g_k, \gamma_1 \dots \gamma_n}  } &=& -\frac{\partial^3 V}{(\partial \phi^i_{g_i, \alpha_1 \dots \alpha_n}) (\partial \phi^j_{g_j, \beta_1 \dots \beta_n} ) (\partial \phi^k_{g_k, \gamma_1 \dots \gamma_n})} \,,\\
{\cal M_S}^{\phi^i_{g_i, \alpha_1 \dots \alpha_n} \phi^j_{g_j, \beta_1 \dots \beta_n} } &=& -\frac{\partial^2 V}{(\partial \phi^i_{g_i, \alpha_1 \dots \alpha_n}) (\partial \phi^j_{g_j, \beta_1 \dots \beta_n} )} \, .
\end{eqnarray}
The objects ${\cal Y}$, ${\cal H}$, ${\cal L}$, ${\cal M}$ and ${\cal M_S}$ are independent of the ordering of their arguments and contain all necessary information about the involved states in the most explicit way and can therefore be used to build up algorithms to calculate the $\beta$ functions for any given model if the particle content and the potential is provided. For this purpose, it is, of course, necessary to express the general formulae in the literature by using these objects. However, this translation is straightforward. We show this at the example of the one-loop $\beta$ function of the Yukawa couplings which reads in the literature \cite{Machacek:1983fi} 
\begin{eqnarray}
\beta_I^a &=& \frac{1}{2} [Y^b Y^{\dagger b} Y^a + Y^a Y^{\dagger b} Y^b] + 2 Y^b Y^{\dagger a} Y^b \nonumber \\
&& + \frac{1}{2} Y^b \mbox{Tr}(Y^{\dagger a} Y^b + Y^{\dagger b} Y^a) - 3 g^2 \{C_2(F), Y^a\} \,.
\end{eqnarray}
Using our most explicit notation, these terms are written as
\begin{eqnarray}
(\beta_I)^{\phi^i_{g_i,\alpha_1 \dots \alpha_n}}_{\psi^j_{g_j, \beta_1 \dots \beta_n} \psi^k_{g_k, \gamma_1 \dots \gamma_n}} &=& \frac{1}{2} 
\sum_{s_1,f_1,f_2}\sum_{o,p,q} \sum_{\eta,\rho,\sigma} \Bigg[{\cal Y}^{\phi^{s_1}_{g_o,\eta_1 \dots \eta_n}}_{\psi^j_{g_j, \beta_1 \dots \beta_n} \psi^{f_1}_{g_p, \rho_1 \dots \rho_n}} \left({\cal Y}^{\phi^{s_1}_{g_o,\eta_1 \dots \eta_n}}_{\psi^{f_1}_{g_p, \rho_1 \dots \rho_n} \psi^{f_2}_{g_q, \sigma_1 \dots \sigma_n}}\right)^\dagger {\cal Y}^{\phi^i_{g_i,\alpha_1 \dots \alpha_n}}_{\psi^{f_2}_{g_q, \sigma_1 \dots \sigma_n} \psi^k_{g_k, \gamma_1 \dots \gamma_n} } + \nonumber \\ 
&& \hspace{1cm}  +  {\cal Y}^{\phi^i_{g_i,\alpha_1 \dots \alpha_n}}_{\psi^j_{g_j, \beta_1 \dots \beta_n} \psi^{f_1}_{g_p, \rho_1 \dots \rho_n}} \left({\cal Y}^{\phi^{s_1}_{g_o,\eta_1 \dots \eta_n}}_{\psi^{f_1}_{g_p, \rho_1 \dots \rho_n} \psi^{f_2}_{g_q, \sigma_1 \dots \sigma_n}}\right)^\dagger {\cal Y}^{\phi^{s_1}_{g_o,\eta_1 \dots \eta_n}}_{\psi^{f_2}_{g_q, \sigma_1 \dots \sigma_n} \psi^k_{g_k, \gamma_1 \dots \gamma_n} }
\Bigg] \nonumber \\
&& +  2 
\sum_{s_1,f_1,f_2}\sum_{o,p,q} \sum_{\eta,\rho,\sigma} \Bigg[{\cal Y}^{\phi^{s_1}_{g_o,\eta_1 \dots \eta_n}}_{\psi^j_{g_j, \beta_1 \dots \beta_n} \psi^{f_1}_{g_p, \rho_1 \dots \rho_n}} \left({\cal Y}^{\phi^i_{g_i,\alpha_1 \dots \alpha_n}}_{\psi^{f_1}_{g_p, \rho_1 \dots \rho_n} \psi^{f_2}_{g_q, \sigma_1 \dots \sigma_n}}\right)^\dagger {\cal Y}^{\phi^{s_1}_{g_o,\eta_1 \dots \eta_n}}_{\psi^{f_2}_{g_q, \sigma_1 \dots \sigma_n} \psi^k_{g_k, \gamma_1 \dots \gamma_n} } \nonumber \\
&&+ \frac{1}{2} 
\sum_{s_1,f_1,f_2}\sum_{o,p,q} \sum_{\eta,\rho,\sigma} \Bigg[ 
{\cal Y}^{\phi^{s_1}_{g_o,\eta_1 \dots \eta_n}}_{\psi^j_{g_j, \beta_1 \dots \beta_n} \psi^k_{g_k, \gamma_1 \dots \gamma_n} } \Bigg( \left({\cal Y}^{\phi^i_{g_i,\alpha_1 \dots \alpha_n} }_{\psi^{f_1}_{g_p, \rho_1 \dots \rho_n} \psi^{f_2}_{g_q, \sigma_1 \dots \sigma_n}}\right)^\dagger {\cal Y}^{\phi^{s_1}_{g_o,\eta_1 \dots \eta_n}}_{\psi^{f_2}_{g_q, \sigma_1 \dots \sigma_n} \psi^{f_1}_{g_p, \rho_1 \dots \rho_n} }
\nonumber \\
&& \hspace{1cm}  + \left({\cal Y}^{\phi^{s_1}_{g_o,\eta_1 \dots \eta_n}}_{\psi^{f_1}_{g_p, \rho_1 \dots \rho_n} \psi^{f_2}_{g_q, \sigma_1 \dots \sigma_n}}\right)^\dagger {\cal Y}^{\phi^i_{g_i,\alpha_1 \dots \alpha_n} }_{\psi^{f_2}_{g_q, \sigma_1 \dots \sigma_n} \psi^{f_1}_{g_p, \rho_1 \dots \rho_n}}  \Bigg) \Bigg] + \nonumber \\
&& - 3 \sum_n g_n^2 \sum_{f_1} \sum_o \sum_\eta \Bigg[C_{2,n}^{\psi^j_{g_j, \beta_1 \dots \beta_n} \psi^{f_1}_{g_o, \eta_1 \dots \eta_n} } {\cal Y}^{\phi^i_{g_i,\alpha_1 \dots \alpha_n}}_{\psi^{f_1}_{g_o, \eta_1 \dots \eta_n} \psi^k_{g_k, \gamma_1 \dots \gamma_n}} \nonumber \\
&& \hspace{5cm}  +  {\cal Y}^{\phi^i_{g_i,\alpha_1 \dots \alpha_n}}_{\psi^j_{g_j, \beta_1 \dots \beta_n} \psi^{f_1}_{g_o, \eta_1 \dots \eta_n}}  C_{2,n}^{ \psi^{f_1}_{g_o, \eta_1 \dots \eta_n} \psi^k_{g_k, \gamma_1 \dots \gamma_n}} \Bigg] \nonumber \\
&& - 3 g^2 \sum_{f_1} \sum_o \sum_\eta \Bigg[C_{2,U(1)}^{\psi^j_{g_j, \beta_1 \dots \beta_n} \psi^{f_1}_{g_o, \eta_1 \dots \eta_n} } {\cal Y}^{\phi^i_{g_i,\alpha_1 \dots \alpha_n}}_{\psi^{f_1}_{g_o, \eta_1 \dots \eta_n} \psi^k_{g_k, \gamma_1 \dots \gamma_n}} \nonumber \\
&& \hspace{5cm}  +  {\cal Y}^{\phi^i_{g_i,\alpha_1 \dots \alpha_n}}_{\psi^j_{g_j, \beta_1 \dots \beta_n} \psi^{f_1}_{g_o, \eta_1 \dots \eta_n}}  C_{2,U(1)}^{ \psi^{f_1}_{g_o, \eta_1 \dots \eta_n} \psi^k_{g_k, \gamma_1 \dots \gamma_n}} \Bigg] \, .
\label{eq:betaYexplicit}
\end{eqnarray} 

The most general expressions look quite involved. Therefore, we are going to clarify their usage at the example of the SM, but concentrate
again on the non-Abelian sector.  The Yukawa part of the SM potential is usually written as:
\begin{equation}
\label{eq:SMyukawa}
-V =  Y^{ij}_d H^\dagger d_i q_j + Y_e^{ij} H^\dagger e_i l_j + Y_u^{ij} H u_i q_j \, .
\end{equation}
Here, $i,j$ are the generation indices of the SM fermions and all isospin and charge indices are only implicit. 
We can make the following association: 
\begin{eqnarray}
&\psi^1_{g_j,\alpha_1 \alpha_2} = q_{j, \alpha_1 \alpha_2} \,, \hspace{1cm}
\psi^2_{g_j,\alpha_1 \alpha_2} = u_{j, \alpha_2} \,,&\\
&\psi^3_{g_j,\alpha_1 \alpha_2} = d_{j, \alpha_2} \,, \hspace{1cm}
\psi^4_{g_j,\alpha_1 \alpha_2} = l_{j, \alpha_1 } \,, \hspace{1cm}
\psi^5_{g_j,\alpha_1 \alpha_2} =  e_{j} \,& \\
& \phi^1_{g_j,\alpha_1 \alpha_2} = \phi^h_{\alpha_1} \,, \hspace{1cm}
\phi^2_{g_j,\alpha_1 \alpha_2} = \sigma^h_{\alpha_1} \, .&
\end{eqnarray}
Together with \ref{eq:decompositioHiggs} and \ref{eq:GeneralPotential}, the potential
given in \ref{eq:SMyukawa} becomes
\begin{eqnarray}
-V &=& \frac{1}{\sqrt{2}} Y^{\phi^h_{\alpha_1}}_{d_{i,\beta_2} q_{j,\gamma_1 \gamma_2}} \delta_{\beta_2 \gamma_2} \delta_{\alpha_1\gamma_1} \phi^h_{\alpha_1} d_{i,\beta_2} q_{j,\gamma_1 \gamma_2} - \frac{i}{\sqrt{2}} Y^{\sigma^h_{\alpha_1}}_{d_{i,\beta_2} q_{j,\gamma_1 \gamma_2}} \delta_{\beta_2 \gamma_2} \delta_{\alpha_1\gamma_1} \sigma^h_{\alpha_1} d_{i,\beta_2} q_{j,\gamma_1 \gamma_2} \nonumber \\
&& + \frac{1}{\sqrt{2}} Y^{\phi^h_{\alpha_1}}_{e_{i} l_{j,\gamma_1}} \delta_{\alpha_1\gamma_1} \phi^h_{\alpha_1} e_{i} l_{j,\gamma_1} - 
\frac{i}{\sqrt{2}} Y^{\sigma^h_{\alpha_1}}_{e_{i} l_{j,\gamma_1}} \delta_{\alpha_1\gamma_1} \sigma^h_{\alpha_1} e_{i} l_{j,\gamma_1} \nonumber \\
&& +\frac{1}{\sqrt{2}} Y^{\phi^h_{\alpha_1}}_{u_{i,\beta_2} q_{j,\gamma_1 \gamma_2}} \delta_{\beta_2 \gamma_2} \epsilon_{\alpha_1\gamma_1} \phi^h_{\alpha_1} u_{i,\beta_2} q_{j,\gamma_1 \gamma_2} + 
\frac{i}{\sqrt{2}} Y^{\sigma^h_{\alpha_1}}_{u_{i,\beta_2} q_{j,\gamma_1 \gamma_2}} \delta_{\beta_2 \gamma_2} \epsilon_{\alpha_1\gamma_1} \sigma^h_{\alpha_1} u_{i,\beta_2} q_{j,\gamma_1 \gamma_2}  \,,
\end{eqnarray}
Here, we already introduced the CGC 
\begin{equation}
C_{\alpha \beta}^{2,2} = \epsilon_{\alpha\beta} \,,\hspace{1cm} C_{\alpha \beta}^{2^*,2} = C_{\alpha \beta}^{2,2^*} = \delta_{\alpha\beta} \,,
\end{equation}
for $\SU{2}$ as well as 
\begin{equation}
C_{\alpha \beta}^{\bar{3},3} = C_{\alpha \beta}^{3,\bar{3}} = \delta_{\alpha\beta} 
\end{equation}
for $\SU{3}$. 
Using \ref{eq:defY} we can calculate the ${\cal Y}$'s we need:
\begin{eqnarray}
&{\cal Y}^{\sigma^h_{\alpha_1}}_{u_{i,\beta_2} q_{j,\gamma_1 \gamma_2}} = \frac{i}{\sqrt{2}} Y_u^{ij} \delta_{\beta_2 \gamma_2} \epsilon_{\alpha_1 \gamma_1} \,, \hspace{1cm}
{\cal Y}^{\phi^h_{\alpha_1}}_{u_{i,\beta_2} q_{j,\gamma_1 \gamma_2}} = \frac{1}{\sqrt{2}} Y_u^{ij} \delta_{\beta_2 \gamma_2} \epsilon_{\alpha_1 \gamma_1}& \\ 
&{\cal Y}^{\sigma^h_{\alpha_1}}_{d_{i,\beta_2} q_{j,\gamma_1 \gamma_2}} = -\frac{i}{\sqrt{2}} Y_d^{ij} \delta_{\beta_2 \gamma_2} \delta_{\alpha_1 \gamma_1} \,, \hspace{1cm}
{\cal Y}^{\phi^h_{\alpha_1}}_{d_{i,\beta_2} q_{j,\gamma_1 \gamma_2}} = \frac{1}{\sqrt{2}} Y_d^{ij} \delta_{\beta_2 \gamma_2} \delta_{\alpha_1 \gamma_1}& \\ 
& {\cal Y}^{\sigma^h_{\alpha_1}}_{e_{i} l_{j,\gamma_1}} =-\frac{i}{\sqrt{2}} Y_e^{ij}  \delta_{\alpha_1 \gamma_1} \,, \hspace{1cm}
{\cal Y}^{\phi^h_{\alpha_1}}_{e_{i} l_{j,\gamma_1}} = \frac{1}{\sqrt{2}} Y_e^{ij}  \delta_{\alpha_1 \gamma_1} \, .&
\end{eqnarray}
All other combinations of fields vanish. Inserting this into \ref{eq:betaYexplicit} and evaluating all sums we would obtain the one-loop $\beta$ function for all Yukawa couplings. For instance, the $\beta$-function of $Y_d$ can be calculated using the relation
\begin{equation}
\beta^I_{Y^{ij}_d} = \sqrt{2} (\beta_I)^{\phi^h_{\alpha_1}}_{d_{i, \beta_2} q_{j, \gamma_1 \gamma_2}} \delta_{\beta_2 \gamma_2} \delta_{\alpha_1 \gamma_1} \equiv \sqrt{2} (\beta_I)^{\phi^h_{1}}_{d_{i,2} q_{j, 1 2}}
\end{equation}
First, we multiplied the $\beta$ function by $\sqrt{2}$ since ${\cal Y}^{\phi^h}_{d q}$ corresponds to $\frac{Y_d}{\sqrt{2}}$, while we want to have the running of $Y_d$. Furthermore, we restricted ourselves to an explicit combination of external color charges and isospin indices. This has been done to simplify the following calculation. To point out the main steps of the calculation, we concentrate on the fourth and fifth line of  \ref{eq:betaYexplicit}:
\begin{eqnarray}
\beta^I_{Y^{ij}_d} = \sqrt{2}  \Bigg[ \dots  \nonumber \\
&& \frac{1}{2} \sum_{s} \sum_{f_1,f_2} \sum_{p,q =1}^{{\cal N}^f_F} \sum_{\eta_1=1}^2 \sum_{\sigma_1=1}^2 \sum_{\rho_1=1}^2 \sum_{\rho_2=1}^3 \sum_{\sigma_2=1}^3 \Bigg[ 
{\cal Y}^{\phi^s_{\eta_1}}_{d_{i,1} q_{j,12} } \Bigg( \left({\cal Y}^{\phi^h_{1} }_{\psi^{f_1}_{p, \rho_1 \rho_2} \psi^{f_2}_{q, \sigma_1 \sigma_2}}\right)^\dagger {\cal Y}^{\phi^{s}_{\eta_1}}_{\psi^{f_2}_{q, \sigma_1 \sigma_2} \psi^{f_1}_{p, \rho_1 \rho_2} } + \nonumber \\
&& \hspace{2cm} +  \left({\cal Y}^{\phi^{s}_{\eta_1}}_{\psi^{f_1}_{p, \rho_1 \rho_2} \psi^{f_2}_{q, \sigma_1 \sigma_2}}\right)^\dagger {\cal Y}^{\phi^h_{1} }_{\psi^{f_2}_{q, \sigma_1 \sigma_2} \psi^{f_1}_{p, \rho_1 \rho_2} } \Bigg) \Bigg] +
\nonumber \\
\dots \Bigg] \, .
\end{eqnarray}
First, one has to evaluate the sum over $s$ and $f_1,f_2$. The non-vanishing contributions are 
\begin{equation}
\psi^{f_{1,2}} = d,q \,, \hspace{1cm} \psi^{f_{1,2}} = u,q  \,, \hspace{1cm} \psi^{f_{1,2}} = l,e
\end{equation}
while only 
\begin{equation}
\phi^s= \phi^h
\end{equation}
is possible. For $\phi^s=\sigma^h$ the two terms in the sum enter with a different sign and cancel each other. All terms are calculated in the same way, and we pick for further discussion $\psi^{f_1} = d$, $\psi^{f_2} = q$:
\begin{eqnarray}
\beta^I_{Y^{ij}_d} &=& \sqrt{2}  \Bigg[ \dots  +\frac{1}{2} 
\sum_{p,q} \sum_{\eta_1,\sigma_1} \sum_{\rho_2,\sigma_2} \Bigg[ 
{\cal Y}^{\phi^h_{\eta_1}}_{d_{i,2} q_{j,12} } \Bigg( \left({\cal Y}^{\phi^h_{1} }_{{d}_{p, \rho_2} q_{q, \sigma_1 \sigma_2}}\right)^\dagger {\cal Y}^{\phi^h_{\eta_1}}_{q_{q, \sigma_1 \sigma_2} d_{p, \rho_2} } + \left({\cal Y}^{\phi^h_{\eta_1}}_{{d}_{p, \rho_2} q_{q, \sigma_1 \sigma_2}}\right)^\dagger {\cal Y}^{\phi^h_{1} }_{q_{q, \sigma_1 \sigma_2} d_{p, \rho_2} }\Bigg) \Bigg]+\dots
\nonumber \\
&& = \dots \frac{\sqrt{2}}{2} \sum_{p,q} \sum_{\eta_1,\sigma_1} \sum_{\rho_2,\sigma_2} \Bigg[ \left(\frac{1}{\sqrt{2}}Y^{ij}_d \delta_{1 \eta_1}\right) \left( \frac{1}{\sqrt{2}}Y^{pq,\dagger}_d \delta_{1 \sigma_1} \delta_{\rho_2 \sigma_2}\right) \left(\frac{1}{\sqrt{2}}Y^{qp}_d  \delta_{\eta_1 \sigma_1} \delta_{\rho_2 \sigma_2} \right) + \nonumber \\
&& \hspace{2cm} + \left(\frac{1}{\sqrt{2}}Y^{ij}_d \delta_{1 \eta_1}\right) \left( \frac{1}{\sqrt{2}}Y^{pq,\dagger}_d \delta_{\eta_1 \sigma_1} \delta_{\rho_2 \sigma_2}\right) \left(\frac{1}{\sqrt{2}}Y^{qp}_d \delta_{1 \sigma_1} \delta_{\rho_2 \sigma_2}\right)  \Bigg] + \dots \nonumber \\
&& = \dots +\,\frac{1}{4} Y^{ij}_d \sum_{p,q} (3 Y^{pq,\dagger}_d Y^{qp}_d + 3 Y^{pq,\dagger}_d Y^{qp}_d) \,+ \dots \nonumber \\
&& = \dots +\,\frac{3}{2} Y^{ij}_d \mbox{Tr}(Y^{\dagger}_d Y_d ) \,+ \dots 
\end{eqnarray}
Here, one can see nicely the appearance of the color factor due to the sum over the charges of the internal particles. The other terms can be obtained similarly: $\psi^{f_1}=q$ and $\psi^{f_2}=d$ results in the same coefficient, i.e. one gains a factor of two. For $\psi^{f_{1,2}} = q,u$ one gets the same result as for $q,d$ with $Y_d$ replaced by $Y_u$, while for $\psi^{f_{1,2}} = l,e$ one gets this term with $Y_d \to Y_e$ together with a relative factor of $\frac{1}{3}$ because of the missing color factor. In sum, we end up with the well known result
\begin{equation}
\beta^I_{Y_d} = \frac{1}{16 \pi^2} \left[Y_d \left(3 \mbox{Tr}(Y^{\dagger}_d Y_d )+3 \mbox{Tr}(Y^{\dagger}_u Y_u )+\mbox{Tr}(Y^{\dagger}_e Y_e )\right) + \dots \right] \, .
\end{equation}
The same approach holds for all other terms and even at the two-loop level. To cover also the evaluation of the quartic coupling and the Higgs mass terms given by
\begin{equation}
-V =  -\frac{1}{2} \lambda |H^\dagger H|^2 + \mu^2 H^\dagger H \, ,
\end{equation}
the following objects are needed in addition:
\begin{eqnarray}
{\cal L}^{\phi^h_{\alpha_1} \phi^h_{\beta_1} \phi^h_{\gamma_1} \phi^h_{\delta_1}   } &=& \lambda ( \delta_{\alpha_1 \beta_1} \delta_{\gamma_1 \delta_1} + \delta_{\alpha_1 \gamma_1} \delta_{\beta_1 \delta_1} + \delta_{\alpha_1 \delta_1} \delta_{\beta_1 \gamma_1}) \\
{\cal L}^{\sigma^h_{\alpha_1} \sigma^h_{\beta_1} \sigma^h_{\gamma_1} \sigma^h_{\delta_1}   } &=& \lambda ( \delta_{\alpha_1 \beta_1} \delta_{\gamma_1 \delta_1} + \delta_{\alpha_1 \gamma_1} \delta_{\beta_1 \delta_1} + \delta_{\alpha_1 \delta_1} \delta_{\beta_1 \gamma_1}) \\
{\cal L}^{\phi^h_{\alpha_1} \phi^h_{\beta_1} \sigma^h_{\gamma_1} \sigma^h_{\delta_1}   } &=& \lambda \delta_{\alpha_1 \beta_1} \delta_{\gamma_1 \delta_1}\\
{\cal M_S}^{\phi^h_{\alpha_1} \phi^h_{\beta_1} } &=& -\mu^2 \delta_{\alpha_1 \beta_1} \\
{\cal M_S}^{\sigma^h_{\alpha_1} \sigma^h_{\beta_1} } &=& -\mu^2 \delta_{\alpha_1 \beta_1} 
\end{eqnarray}
As an example to demonstrate this, we pick the term $~\propto \mu^2 \lambda$ in the one-loop $\beta$-function of $\mu^2$. This terms is stemming from
\begin{equation}
\beta^I_{m_{ab}^2} = m_{ef}^2 \lambda_{abef} + \dots 
\end{equation}
which results in the most explicit form in
\begin{equation}
\beta^I_{\mu^2} = -(\beta^I)^{\phi^h_i \phi^h_j} \equiv  -(\beta^I)^{\phi^h_1 \phi^h_1} = - \sum_{s_1,s_2} \sum_{\eta_1,\rho_1} {\cal M_S}^{\phi^{s_1}_{\eta_1} \phi^{s_2}_{\rho_1} } {\cal L}^{{\phi^{s_1}_{\eta_1} \phi^{s_2}_{\rho_1} } \phi^h_1 \phi^h_1  } + \dots
\end{equation}
Here, we used again a particular choice for the isospin indices. The only, non-vanishing combinations are $\phi^{s_1}=\phi^{s_2} = \sigma^h$ and $\phi^{s_1}=\phi^{s_2}=\phi^h$. Hence, we obtain 
\begin{eqnarray}
\beta^I_{\mu^2} &=& - \sum_{\eta_1,\rho_1} \left[ {\cal M_S}^{\sigma^h_{\eta_1} \sigma^h_{\rho_1} } {\cal L}^{{\sigma^h_{\eta_1} \sigma^h_{\rho_1} } \phi^h_1 \phi^h_1   } + {\cal M_S}^{\phi^h_{\eta_1} \phi^h_{\rho_1} } {\cal L}^{{\phi^h_{\eta_1} \phi^h_{\rho_1} } \phi^h_1 \phi^h_1   } \right] + \dots \nonumber \\
&& = - \sum_{\eta_1,\rho_1}\left[-\mu^2  \delta_{\eta_1\rho_1} \lambda \delta_{\eta_1 \rho_1} - \mu^2 \delta_{\eta_1 \rho_1} \lambda (\delta_{\eta_1 \rho_1} + 2\delta_{1\eta_1} \delta_{1\rho_1} )]\right]+\dots \nonumber \\
&& = 6 \mu^2 \lambda + \dots
\end{eqnarray}

\section{Validation}
\label{sec:validation}

In the literature, there are very few models for which the RGEs at two-loop have been calculated for all dimensionless and dimensionful parameters. Therefore, we have also independently developed routines in {\tt Mathematica} to calculate the 
full two-loop RGEs for all terms. These routines will be merged with the \sarah \cite{Staub:2008uz,Staub:2009bi,Staub:2010jh,Staub:2012pb} in an upcoming version. For all tested models we had full agreement between \pyrate and the results obtained by the new \sarah routines.

In the following, we present the comparisons between the results obtained with \pyrate and the RGEs for some models presented in the literature. The reason for choosing this subset of models is twofold. For one thing, they represent a broad variety of interactions so that we could obtain non-trivial tests for \pyrate. For another, these are the models for which the RGEs have been calculated at two-loop for a large number of parameters.

\subsubsection*{Standard Model}
We find full agreement for all parameters at the two-loop level 
with the results given in Ref.~\cite{Luo:2002ey} including the full CP and flavor 
structure. This also confirms that the differences pointed out in Ref.~\cite{Luo:2002ey}
in comparison to the earlier results of Refs.~\cite{Machacek:1983fi,Machacek:1984zw} are correct. 
\subsubsection*{Standard Model with real scalar singlet}
For the SM extended by a real scalar singlet field and a Dirac doublet \cite{Cheung:2012nb} we find complete agreement for all dimensionless parameters, where we have applied the same approximation that only third generation
Yukawa couplings contribute. 
\subsubsection*{Standard Model with real scalar triplet}
The RGEs for all dimensionless parameters for the SM extended by 
a real scalar triplet and a Dirac doublet are given in Ref.~\cite{Cheung:2012nb}. Making the same approximation as in Ref.~\cite{Cheung:2012nb}, i.e.~neglecting the Yukawa interactions 
for the first two generations of SM fermions, we find disagreement at the one- and two-loop level in the following parameters: $\Delta b^{(2)}_{\lambda_H}$, $b^{(1)}_{\kappa_T}$, $b^{(2)}_{\kappa_T}$, $b^{(1)}_{\lambda_T}$, $b^{(2)}_{\lambda_T}$.
\subsubsection*{Standard Model with Majorana singlet fermion and Dirac doublet}
The RGEs for all dimensionless parameters for the SM extended by 
a real singlet fermion and a Dirac doublet are given in Ref.~\cite{Cheung:2012nb}. We find complete
agreement with these results by taking the same approximation that only third generation
Yukawa couplings contribute. 
\subsubsection*{Standard Model with complex scalar doublet}
The two Higgs doublet model is one of the most widely studied extensions of the SM. In the literature, the results for 
the $\beta$-functions are mostly available at one-loop level, see e.g.~Ref.~\cite{Branco:2011iw} and references therein. We agree with those results. In addition, 
Ref.~\cite{Cheung:2012nb} contains also partial two-loop results for which we also find agreement in the limit that only third generation Yukawas are taken into account. 
\subsubsection*{Standard Model with Majorana triplet fermion and Dirac doublet}
The one- and two-loop $\beta$-functions for all dimensionless parameters for the 
SM extended by a fermionic Majorana triplet and Dirac doublet are given in Ref.~\cite{Cheung:2012nb}. We find full agreement with their results, if we
\begin{inparaenum}[(i)]
\item take the limit of vanishing Yukawa couplings for the first two generations,
\item include a relative factor of $\sqrt{2}$ in the definition of the Yukawa-like couplings of the triplet. 
\end{inparaenum}
This factor stems from a different normalization of the triplet (see also \ref{sec:pitfalls}).
\subsubsection*{$B-L$ extended Standard Model}
The one-loop RGEs for an extension of the SM by an additional $\U{1}_{B-L}$, right-handed neutrinos and a SM singlet complex scalar with $B-L$ charge 2 have been calculated in Ref.~\cite{Basso:2010jm}. These results contain kinetic mixing that at present is not calculated by \pyrate yet. In the limit of $\tilde{g} \to 0$
we find almost full agreement. Only the coefficient of the contribution proportional
to Tr($Y_M^4$) in the RGE for $\lambda_2$ should read -16 and not -1. This issue has been confirmed 
in a private discussion with one of the authors of Ref.~\cite{Basso:2010jm}. 
\subsubsection*{SM extended by a complex triplet and vectorlike doublets}
Partial one-loop results for the SM extended by a complex scalar and vectorlike doublets are 
given in Ref.~\cite{Arina:2012fb}. 
However, we find the following disagreements:
In the quartic coupling $\lambda_{\Delta H}$ the terms Tr($f_L^\dagger f_L f_L^\dagger f_L + f_\psi^\dagger f_\psi f_\psi^\dagger f_\psi$)
cannot be present, since they would need four triplets as external fields. In the $\beta$-function for the Yukawa couplings $f_L$ and $f_\psi$, we do not find the terms $3 f_L f_\psi^\dagger f_\psi$ and $3 f_\psi f_L^\dagger f_L$, respectively. In addition, we also find disagreement in the coefficients for the trilinear coupling. The full details can be obtained by running \pyrate. 
\subsubsection*{SM with neutrino Yukawa couplings}
The RGEs for the SM extended by right handed neutrinos have been given at the two-loop level for all dimensionless parameters in Ref.~\cite{Pirogov:1998tj}. However, as it was already pointed out in Ref.~\cite{Wingerter:2011dk}, the terms at two-loop are missing. In addition, we find many more terms in disagreement with Ref.~\cite{Pirogov:1998tj}. Also, we have some disagreement in some terms
in the two-loop $\beta$-function of $\lambda$. 

\subsubsection*{SM with a fourth generation of vectorlike fermions}
The SM extended by a vector-like fourth generation
has been studied in Ref.~\cite{Ishiwata:2011hr}. 
The authors have calculated the RGEs for the Yukawa couplings and the quartic Higgs coupling at one-loop in the limit, where the fourth generation quark masses are of the order the cut-off scale. The results we obtain with \pyrate are identical for all the couplings they have listed.

%
\section{Conclusions}
\label{sec:conclusion}

To the present date, the automated generation of two-loop renormalization group equations was available only for supersymmetric models. In the present article, we attempted to close this gap and introduced \pyrate that automatically generates for a general gauge theory the two-loop RGEs for all dimensionless and dimensionful parameters. \pyrate is easy to use: Once the user specifies in an intuitive format the gauge group and the particle content of a given model, the RGEs are generated, and, for ease of inspection, directly exported to \LaTeX{}. Also, the results can be exported to \mathematica where the RGEs can be numerically solved and plotted. Furthermore, we have developed an  interactive mode in form of an \code{IPython Notebook} that mimics much of the functionality of \mathematica.

Since the calculations that lead to the RGEs are if not difficult so at least involved, we paid special attention to validating our results. To that end, we not only compared the RGEs generated by \pyrate with complete or partial results that are available in the literature, but also developed \mathematica routines that will be part of an upcoming version of \sarah 4. With \sarah 4 we find complete agreement, whereas we have some disagreement with the literature, as elaborated on in the previous section. 

In future we plan to extend the functionality of \pyrate to 
\begin{inparaenum}[(i)]
\item include kinetic mixing,
\item include partial 3-loop contributions to the RGEs,
\item extend the library of gauge groups and irreps,
\item support generation indices for scalars, and
\item run the VEVs (including their gauge dependence).
\end{inparaenum}

We believe that \pyrate can make an important contribution to exploring physics beyond the Standard Model. We have developed the code in the spirit that calculational or technical details should not stop us exploring new scenarios and that one should make sensible use of computer-aided calculations. We hope that the high-energy physics community will find PyR@TE
useful and we encourage interested readers to send us constructive feedback
which will be helpful to further improve future versions.

\section*{Acknowledgments}

We would like to thank Lorenzo Basso, Renato Fonseca, D.R.T.~Jones, Werner Porod, and Stuart Raby for useful discussions. We are also indebted to Tom\'a\v{s} Je\v{z}o and Josselin Proudom for carefully proof-reading the draft of this article. F.L.~would like to thank Tom\'a\v{s} Je\v{z}o for introducing him to Python and for always patiently answering his questions, and  Matthew Rocklin for clarifying questions regarding the backward compatibility of SymPy. We are indebted to HepForge for hosting our project on their web pages. This work has been
supported by a Ph.D. fellowship of the French Ministry
for Education and Research and by the Theory-LHC-
France initiative of the CNRS/IN2P3. 

\appendix
\clearpage
\newpage
\labelformat{section}{#1}

\section{List of available irreducible representations}
\label{app:irreps}
We list below all the gauge groups with their respective irreps available in \pyrate.

\begin{table}[!h]
\caption{List of all the irreps available in {\pyrate}. Note that the True argument for $\SU{2}$ represents the conjugate representation.}
\label{tab:irreps}
\vspace*{0.3cm}
\center
	\begin{tabular}{|l|l|l|l|}
	\hline
		Gauge Group& Irreps: dimension &Gauge Group& Irreps: dimension\\\hline\hline
		$\SU{2}$&(0,) : 1&$\SU{4}$& (0,0,0) : 1\\
		 &(1,) : 2 &&(0,0,1) : 4\\
		 &(1,True) : 2&&(0,0,2) : 10\\
		 &(2,) : 3&&(0,1,0) : 6 \\
		 &(2,True) : 3&&(1,0,0) : 4\\
		 &(3,) : 4	 &&(1,0,1) : 15\\
		 &(3,True) : 4&&(2,0,0) : 10\\\hline
		 $\SU{3}$& (0,0) : 1&$\SU{5}$&(0,0,0,0) : 1\\
		 &(0,1) : 3&&(0,0,0,1) : 5\\
		 &(0,2) : 6&&(0,0,0,2) : 15\\
		 &(0,3) : 10&&(0,0,1,0) : 10\\
		 &(1,0) : 3&&(0,1,0,0) : 10\\
		 &(1,1) : 8&&(1,0,0,0) : 5\\
		 &(2,0) : 6&&(1,0,0,1) : 24\\
		 &(3,0) : 10&&(2,0,0,0) : 15\\\hline 
		 $\SU{6}$ & (0,0,0,0,0) : 1&$\U{1}_Y$&\\
		 &(0,0,0,0,1) : 6&&\\
		 &(0,0,0,0,2) : 21 &&\\
		 &(0,0,0,1,0) : 15 &&\\
		 &(0,0,1,0,0) : 20 &&\\
		 &(0,1,0,0,0) : 15 && \\
		 &(1,0,0,0,0) : 6 &&\\
		 &(1,0,0,0,1) : 35 &&\\
		 &(2,0,0,0,0) : 21&&\\\hline
		 
	\end{tabular}
\end{table}

\newpage
\section{Sample Model Files}
\label{app:sample_model_files}
\begin{lstlisting}[basicstyle=\LSTfont, label=lst:SMBiD,caption=models/SM\_BiD.model]
# YAML 1.1
---
Author: Florian Lyonnet
Date: 9.08.2013
Name: SMBiD
Groups: {'U1': U1, 'SU2L': SU2, 'SU2R': SU2}

##############################
#Fermions assumed weyl spinors
##############################
Fermions: {
   QL: {Gen: 3, Qnb:{ U1: 1/6, SU2L: 2, SU2R: 1}},
   QR: {Gen: 3, Qnb:{ U1: -1/6, SU2L: 1, SU2R: 2}},
   LL: {Gen: 3, Qnb:{ U1: -1/2, SU2L: 2, SU2R: 1}},
   LR: {Gen: 3, Qnb:{ U1: 1/2, SU2L: 1, SU2R: 2}},
}

#############
#Real Scalars
#############

RealScalars: {
  Pi: {U1: 0, SU2L: 2, SU2R: 2},
  Sigma: {U1: 0, SU2L: 2, SU2R: 2},
}

##########################################################################
#Complex Scalars : have to be expressed in terms of Real Scalars see above
##########################################################################

CplxScalars: {
  H: {RealFields: [Pi,I*Sigma], Norm: 1/Sqrt(2), Qnb : {U1: 0, SU2L: 2, SU2R: 2}},
  H*: {RealFields: [Pi,-I*Sigma], Norm: 1/Sqrt(2), Qnb : {U1: 0, SU2L: -2, SU2R: -2}}
}

Potential: {

#######################################
# All particles must be defined above !
#######################################
 
Yukawas:{
 'Y_{q}': {Fields: [H,QL,QR], Norm: 1}, 
 'Y_{l}': {Fields: [H,LL,LR], Norm: 1} 
 },
QuarticTerms: {
'\lambda_{1}' : {Fields : [H,H*,H,H*], Norm : 1/2}
 },
ScalarMasses: {
'\mu_{1}' : {Fields : [H*,H], Norm : 1}
 }
}
\end{lstlisting}

\newpage
\begin{lstlisting}[basicstyle=\LSTfont, label=lst:SMCplexDoubletScalar,caption=models/SMCplexDoubletScalar.model]
# YAML 1.1
# #This is the A.4 Model of 1203.5106
---
Author: Florian Lyonnet
Date: 26.07.2013
Name: SMCplexDoubletScalar
Groups: {'U1': U1, 'SU2L': SU2, 'SU3c': SU3}

##############################
#Fermions assumed weyl spinors
##############################
Fermions: {
   Qbar: {Gen: 3, Qnb:{ U1: -1/6, SU2L: -2, SU3c: -3}},
   Lbar: {Gen: 3, Qnb:{ U1: 1/2, SU2L: -2, SU3c: 1}},
   uR: {Gen: 3, Qnb:{ U1: 2/3, SU2L: 1, SU3c: [1,0]}},
   dR: {Gen: 3, Qnb:{ U1: -1/3, SU2L: 1, SU3c: 3}},
   eR: {Gen: 3, Qnb:{ U1: -1, SU2L: 1, SU3c: 1}},
}

#############
#Real Scalars
#############

RealScalars: {
  Pi: {U1: 1/2, SU2L: 2, SU3c: 1},
  Sigma: {U1: 1/2, SU2L: 2, SU3c: 1},
  PiD : {U1: 1/2, SU2L: 2, SU3c: 1},
  SigmaD : {U1: 1/2, SU2L: 2, SU3c: 1}
}

##########################################################################
#Complex Scalars : have to be expressed in terms of Real Scalars see above
##########################################################################

CplxScalars: {
  H: {RealFields: [Pi,I*Sigma], Norm: 1/Sqrt(2), Qnb : {U1: 1/2, SU2L: 2, SU3c: 1}},
  H*: {RealFields: [Pi,-I*Sigma], Norm: 1/Sqrt(2), Qnb : {U1: -1/2, SU2L: -2, SU3c: 1}},
  D: {RealFields: [PiD,I*SigmaD], Norm: 1/Sqrt(2), Qnb : {U1: 1/2, SU2L: 2, SU3c: 1}},
  D*: {RealFields: [PiD,-I*SigmaD], Norm: 1/Sqrt(2), Qnb : {U1: -1/2, SU2L: -2, SU3c: 1}},
}

Potential: {

#######################################
# All particles must be defined above !
#######################################
 
Yukawas:{
  'Y_{u}': {Fields: [Qbar,uR,H*], Norm: 1},
  'Y_{d}': {Fields: [Qbar,dR,H], Norm: 1},
  'Y_{e}': {Fields: [Lbar,eR,H], Norm: 1}
 },
QuarticTerms: {
'\lambda_{1}' : {Fields : [H,H*,H,H*], Norm : 1/2},
'\lambda_{D}' : {Fields: [D,D*,D,D*], Norm : 1/2},
'\kappa_{D}' : {Fields: [D,D*,H,H*], Norm : 1/2},
'\Pkappa_{D}' : {Fields: [D,H*,H,D*], Norm : 1/2}
 },
ScalarMasses: {
'\mu_{H}' : {Fields : [H,H*], Norm : 1},
'\mu_{D}' : {Fields : [D,D*], Norm : 1}
 }
}
\end{lstlisting}

\newpage
\begin{lstlisting}[basicstyle=\LSTfont, label=lst:SM,caption=models/SM.model]
# YAML 1.1
---
Author: Florian Lyonnet
Date: 11.06.2013
Name: SM
Groups: {'U1': U1, 'SU2L': SU2, 'SU3c': SU3}

##############################
#Fermions assumed weyl spinors
##############################
Fermions: {
   Qbar: {Gen: 3, Qnb:{ U1: -1/6, SU2L: -2, SU3c: -3}},
   Lbar: {Gen: 3, Qnb:{ U1: 1/2, SU2L: -2, SU3c: 1}},
   uR: {Gen: 3, Qnb:{ U1: 2/3, SU2L: 1, SU3c: [1,0]}},
   dR: {Gen: 3, Qnb:{ U1: -1/3, SU2L: 1, SU3c: 3}},
   eR: {Gen: 3, Qnb:{ U1: -1, SU2L: 1, SU3c: 1}}
}

#############
#Real Scalars
#############

RealScalars: {
  Pi: {U1: 1/2, SU2L: 2, SU3c: 1},
  Sigma: {U1: 1/2, SU2L: 2, SU3c: 1},
}

##########################################################################
#Complex Scalars : have to be expressed in terms of Real Scalars see above
##########################################################################

CplxScalars: {
  H: {RealFields: [Pi,I*Sigma], Norm: 1/Sqrt(2), Qnb : {U1: 1/2, SU2L: 2, SU3c: 1}},
  H*: {RealFields: [Pi,-I*Sigma], Norm: 1/Sqrt(2), Qnb : {U1: -1/2, SU2L: -2, SU3c: 1}}
}

Potential: {

#######################################
# All particles must be defined above !
#######################################
 
Yukawas:{
  'Y_{u}': {Fields: [Qbar,uR,H*], Norm: 1},
  'Y_{d}': {Fields: [Qbar,dR,H], Norm: 1},
  'Y_{e}': {Fields: [Lbar,eR,H], Norm: 1}
 },
QuarticTerms: {
'\Lambda_{1}' : {Fields : [H,H*,H,H*], Norm : 1/2}
 },
ScalarMasses: {
'\mu_{1}' : {Fields : [H*,H], Norm : 1}
 }
}
\end{lstlisting}

\newpage
\begin{lstlisting}[basicstyle=\LSTfont, label=lst:ScalarSinglet,caption=models/ScalarSinglet.model]
# YAML 1.1
---
Author: Florian Lyonnet
Date: 22.07.2013
Name: ScalarSinglet
Groups: {'U1': U1, 'SU2L': SU2, 'SU3c': SU3}

##############################
#Fermions assumed weyl spinors
##############################
Fermions: {
   Qbar: {Gen: 2, Qnb:{ U1: -1/6, SU2L: -2, SU3c: -3}},
   Lbar: {Gen: 2, Qnb:{ U1: 1/2, SU2L: -2, SU3c: 1}},
   uR: {Gen: 3, Qnb:{ U1: 2/3, SU2L: 1, SU3c: [1,0]}},
   dR: {Gen: 3, Qnb:{ U1: -1/3, SU2L: 1, SU3c: 3}},
   eR: {Gen: 3, Qnb:{ U1: -1, SU2L: 1, SU3c: 1}}
}

############
#Real Scalars
#############

RealScalars: {
  Pi: {U1: 1/2, SU2L: 2, SU3c: 1},
  Sigma: {U1: 1/2, SU2L: 2, SU3c: 1},
  si : {U1: 0, SU2L: 1, SU3c: 1}
}

##########################################################################
#Complex Scalars : have to be expressed in terms of Real Scalars see above
##########################################################################

CplxScalars: {
  H: {RealFields: [Pi,I*Sigma], Norm: 1/Sqrt(2), Qnb : {U1: 1/2, SU2L: 2, SU3c: 1}},
  H*: {RealFields: [Pi,-I*Sigma], Norm: 1/Sqrt(2), Qnb : {U1: -1/2, SU2L: -2, SU3c: 1}}
}

Potential: {

#######################################
# All particles must be defined above !
#######################################
 Yukawas:{
  'Y_{u}': {Fields: [Qbar,uR,H*], Norm: 1},
  'Y_{d}': {Fields: [Qbar,dR,H], Norm: 1},
  'Y_{e}': {Fields: [Lbar,eR,H], Norm: 1}
 },

QuarticTerms: {
'\lambda_{1}' : {Fields : [H,H*,H,H*], Norm: 1/2},
 '\lambda_{s}' : {Fields : [si,si,si,si], Norm: 1/2},
 '\kappa_{s}' : {Fields : [H,H*,si,si], Norm: 1/2}
 },

ScalarMasses: {
 '\mu_{1}' : {Fields: [H,H*], Norm: 1},
 '\mu_{s}' : {Fields: [si,si], Norm: 1/2}
 }
}
\end{lstlisting}

\newpage
\begin{lstlisting}[basicstyle=\LSTfont, label=lst:SMSingletDoublet,caption=models/SMSingletDoublet.model]
# YAML 1.1
---
Author: Florian Lyonnet
Date: 30.07.2013
Name: SMSingletDoublet
Groups: {'U1': U1, 'SU2L': SU2, 'SU3c': SU3}

##############################
#Fermions assumed weyl spinors
##############################
Fermions: {
   Qbar: {Gen: 3, Qnb:{ U1: -1/6, SU2L: -2, SU3c: -3}},
   Lbar: {Gen: 3, Qnb:{ U1: 1/2, SU2L: -2, SU3c: 1}},
   uR: {Gen: 3, Qnb:{ U1: 2/3, SU2L: 1, SU3c: [1,0]}},
   dR: {Gen: 3, Qnb:{ U1: -1/3, SU2L: 1, SU3c: 3}},
   eR: {Gen: 3, Qnb:{ U1: -1, SU2L: 1, SU3c: 1}},
   D: {Gen : 1, Qnb:{ U1:  -1/2, SU2L: 2, SU3c: 1}},
   Dc: {Gen: 1, Qnb:{ U1: 1/2, SU2L: 2,SU3c: 1}},
   S: {Gen: 1, Qnb:{ U1: 0, SU2L: 1, SU3c: 1}}
}

#############
#Real Scalars
#############

RealScalars: {
  Pi: {U1: 1/2, SU2L: 2, SU3c: 1},
  Sigma: {U1: 1/2, SU2L: 2, SU3c: 1},
}

##########################################################################
#Complex Scalars : have to be expressed in terms of Real Scalars see above
##########################################################################

CplxScalars: {
  H: {RealFields: [Pi,I*Sigma], Norm: 1/Sqrt(2), Qnb : {U1: 1/2, SU2L: 2, SU3c: 1}},
  H*: {RealFields: [Pi,-I*Sigma], Norm: 1/Sqrt(2), Qnb : {U1: -1/2, SU2L: -2, SU3c: 1}}
}

Potential: {

#######################################
# All particles must be defined above !
#######################################
 
Yukawas:{
  'Y_{u}': {Fields: [Qbar,uR,H*], Norm: 1},
  'Y_{d}': {Fields: [Qbar,dR,H], Norm: 1},
  'Y_{e}': {Fields: [Lbar,eR,H], Norm: 1},
  'g_{d}': {Fields: [H,S,D], Norm: 1/Sqrt(2)},
  'g_{u}': {Fields: [H*,S,Dc], Norm: 1/Sqrt(2)}
 },
QuarticTerms: {
 '\lambda_1' : {Fields : [H,H*,H,H*], Norm : 1/2}
 },
ScalarMasses: {
 '\mu_1' : {Fields : [H*,H], Norm : 1},
 },
FermionMasses:{
 '\mD': {Fields: [D,Dc], Norm: 1},
 '\mS': {Fields: [S,S], Norm: 1/2}
}
}
\end{lstlisting}

\newpage
\begin{lstlisting}[basicstyle=\LSTfont, label=lst:SMCplxTriplet,caption=models/SMCplxTriplet.model]
# YAML 1.1
---
Author: Florian Lyonnet
Date: 9.07.2013
Name: SMCplxTriplet
Groups: {'U1': U1, 'SU2L': SU2, 'SU3c': SU3}

##############################
#Fermions assumed weyl spinors
##############################
Fermions: {
   Qbar: {Gen: 3, Qnb:{ U1: -1/6, SU2L: -2, SU3c: -3}},
   L: {Gen: 3, Qnb:{ U1: -1/2, SU2L: 2, SU3c: 1}},
   uR: {Gen: 3, Qnb:{ U1: 2/3, SU2L: 1, SU3c: 3}},
   dR: {Gen: 3, Qnb:{ U1: -1/3, SU2L: 1, SU3c: 3}},
   eR: {Gen: 3, Qnb:{ U1: -1, SU2L: 1, SU3c: 1}},
   PsiL: {Gen: 1, Qnb: {U1: -1/2, SU2L: 2, SU3c: 1}},
   PsiRbar: {Gen: 1, Qnb: {U1: 1/2, SU2L: -2, SU3c: 1}}
}

#############
#Real Scalars
#############

RealScalars: {
  Pi: {U1: 1/2, SU2L: 2, SU3c: 1},
  Sigma: {U1: 1/2, SU2L: 2, SU3c: 1},
  T1 :{U1: 1, SU2L: 3, SU3c: 1 },
  T2 :{U1: 1, SU2L: 3, SU3c: 1 }
}

##########################################################################
#Complex Scalars : have to be expressed in terms of Real Scalars see above
##########################################################################

CplxScalars: {
  H: {RealFields: [Pi,I*Sigma], Norm: 1/Sqrt(2), Qnb : {U1: 1/2, SU2L: 2, SU3c: 1}},
  H*: {RealFields: [Pi,-I*Sigma], Norm: 1/Sqrt(2), Qnb : {U1: -1/2, SU2L: -2, SU3c: 1}},
  T : {RealFields: [T1,I*T2], Norm: 1/Sqrt(2), Qnb: {U1: 1, SU2L : 3, SU3c: 1}},
  T* : {RealFields: [T1,-I*T2], Norm: 1/Sqrt(2), Qnb: {U1: -1, SU2L : 3, SU3c: 1}}
}

Potential: {

#######################################
# All particles must be defined above !
#######################################
##############
#The doublet vector like and the yukawa corresponding to the triplet are not included yet 
Yukawas:{
  'Y_{u}': {Fields: [H*,Qbar,uR], Norm: 1},
  'f_{L}': {Fields: [T,L,L], Norm: 1/Sqrt(2)},
	'f_{\psi}': {Fields: [T, PsiL,PsiL], Norm: 1/Sqrt(2)}
 },
QuarticTerms: {
'\lambda_{1}' : {Fields : [H,H*,H,H*], Norm : 1/2},
'\lambda_{T}' : {Fields: [T,T*,T,T*], Norm: 1/2},
'\kappa_{T}': {Fields: [T,T*,H,H*], Norm: 1}
 },
ScalarMasses: {
'\mu_{1}' : {Fields : [H,H*], Norm : 1},
 mT : {Fields: [T,T*], Norm: 1/2},
 },
TrilinearTerms: {
 fH : {Fields: [T*,H,H], Norm: 1/Sqrt(2)},
 },
FermionMasses : {
 mD : {Fields: [PsiL,PsiRbar], Norm: 1, latex: \m_D},
 }
}
		
\end{lstlisting}

\newpage
\begin{lstlisting}[basicstyle=\LSTfont, label=lst:SMTripletDoublet,caption=models/SMTripletDoublet.model]
# YAML 1.1
---
Author: Florian Lyonnet
Date: 30.07.2013
Name: SMTripletDoublet
Groups: {'U1': U1, 'SU2L': SU2, 'SU3c': SU3}

##############################
#Fermions assumed weyl spinors
##############################
Fermions: {
   Qbar: {Gen: 3, Qnb:{ U1: -1/6, SU2L: -2, SU3c: -3}},
   Lbar: {Gen: 3, Qnb:{ U1: 1/2, SU2L: -2, SU3c: 1}},
   uR: {Gen: 3, Qnb:{ U1: 2/3, SU2L: 1, SU3c: [1,0]}},
   dR: {Gen: 3, Qnb:{ U1: -1/3, SU2L: 1, SU3c: 3}},
   eR: {Gen: 3, Qnb:{ U1: -1, SU2L: 1, SU3c: 1}},
   D: {Gen : 1, Qnb:{ U1:  -1/2, SU2L: 2, SU3c: 1}},
   Dc: {Gen: 1, Qnb:{ U1: 1/2, SU2L: 2,SU3c: 1}},
   T: {Gen: 1, Qnb: { U1: 0, SU2L: 3, SU3c: 1}}
}

#############
#Real Scalars
#############

RealScalars: {
  Pi: {U1: 1/2, SU2L: 2, SU3c: 1},
  Sigma: {U1: 1/2, SU2L: 2, SU3c: 1},
}

##########################################################################
#Complex Scalars : have to be expressed in terms of Real Scalars see above
##########################################################################

CplxScalars: {
  H: {RealFields: [Pi,I*Sigma], Norm: 1/Sqrt(2), Qnb : {U1: 1/2, SU2L: 2, SU3c: 1}},
  H*: {RealFields: [Pi,-I*Sigma], Norm: 1/Sqrt(2), Qnb : {U1: -1/2, SU2L: -2, SU3c: 1}}
}

Potential: {

#######################################
# All particles must be defined above !
#######################################
 
Yukawas:{
  'Y_{u}': {Fields: [Qbar,uR,H*], Norm: 1},
  'Y_{d}': {Fields: [Qbar,dR,H], Norm: 1},
  'Y_{e}': {Fields: [Lbar,eR,H], Norm: 1},
  'g_{d}': {Fields: [T,D,H], Norm: -1},
  'g_{u}': {Fields: [T,Dc,H*] ,Norm: 1}
 },
QuarticTerms: {
 '\lambda_1' : {Fields : [H,H*,H,H*], Norm : 1/2},
 },
ScalarMasses: {
 '\mu_1' : {Fields : [H,H*], Norm : 1},
 },
FermionMasses: { 
 'm_{T}' : {Fields: [T,T], Norm: 1/2},
 'm_{D}' : {Fields: [D,Dc], Norm: 1}
}
}

\end{lstlisting}

\newpage
\addcontentsline{toc}{section}{References}
\bibliography{mybibliography}

\providecommand{\href}[2]{#2}\begingroup\raggedright\begin{thebibliography}{10}

\bibitem{Aad:2012tfa}
{\bf ATLAS} Collaboration, G.~Aad {\em et al.}, ``{Observation of a new
  particle in the search for the Standard Model Higgs boson with the ATLAS
  detector at the LHC},'' {\em Phys.Lett.} {\bf B716} (2012) 1--29,
\href{http://www.arXiv.org/abs/1207.7214}{{\tt 1207.7214}}.

\bibitem{Chatrchyan:2012ufa}
{\bf CMS} Collaboration, S.~Chatrchyan {\em et al.}, ``{Observation of a new
  boson at a mass of 125 GeV with the CMS experiment at the LHC},'' {\em
  Phys.Lett.} {\bf B716} (2012) 30--61,
\href{http://www.arXiv.org/abs/1207.7235}{{\tt 1207.7235}}.

\bibitem{Sher:1988mj}
M.~Sher, ``{Electroweak Higgs Potentials and Vacuum Stability},'' {\em
  Phys.Rept.} {\bf 179} (1989)
273--418.

\bibitem{Holthausen:2011aa}
M.~Holthausen, K.~S. Lim, and M.~Lindner, ``{Planck scale Boundary Conditions
  and the Higgs Mass},'' {\em JHEP} {\bf 1202} (2012) 037,
\href{http://www.arXiv.org/abs/1112.2415}{{\tt 1112.2415}}.

\bibitem{Degrassi:2012ry}
G.~Degrassi, S.~Di~Vita, J.~Elias-Miro, J.~R. Espinosa, G.~F. Giudice, {\em et
  al.}, ``{Higgs mass and vacuum stability in the Standard Model at NNLO},''
  {\em JHEP} {\bf 1208} (2012) 098,
\href{http://www.arXiv.org/abs/1205.6497}{{\tt 1205.6497}}.

\bibitem{Nilles:1984ge}
H.~P. Nilles, ``Supersymmetry, supergravity and particle physics,'' {\em Phys.
  Rept.} {\bf 110} (1984)
1.

\bibitem{Haber:1984rc}
H.~E. Haber and G.~L. Kane, ``{The Search for Supersymmetry: Probing Physics
  Beyond the Standard Model},'' {\em Phys.Rept.} {\bf 117} (1985)
75--263.

\bibitem{Martin:1997ns}
S.~P. Martin, ``{A Supersymmetry primer},''
\href{http://www.arXiv.org/abs/hep-ph/9709356}{{\tt hep-ph/9709356}}.

\bibitem{Ellwanger:2006rn}
U.~Ellwanger and C.~Hugonie, ``{NMSPEC: A Fortran code for the sparticle and
  Higgs masses in the NMSSM with GUT scale boundary conditions},'' {\em
  Comput.Phys.Commun.} {\bf 177} (2007) 399--407,
\href{http://www.arXiv.org/abs/hep-ph/0612134}{{\tt hep-ph/0612134}}.

\bibitem{Porod:2003um}
W.~Porod, ``{SPheno, a program for calculating supersymmetric spectra, SUSY
  particle decays and SUSY particle production at e+ e- colliders},'' {\em
  Comput.Phys.Commun.} {\bf 153} (2003) 275--315,
\href{http://www.arXiv.org/abs/hep-ph/0301101}{{\tt hep-ph/0301101}}.

\bibitem{Porod:2011nf}
W.~Porod and F.~Staub, ``{SPheno 3.1: Extensions including flavour, CP-phases
  and models beyond the MSSM},'' {\em Comput.Phys.Commun.} {\bf 183} (2012)
  2458--2469,
\href{http://www.arXiv.org/abs/1104.1573}{{\tt 1104.1573}}.

\bibitem{Allanach:2001kg}
B.~Allanach, ``{SOFTSUSY: a program for calculating supersymmetric spectra},''
  {\em Comput.Phys.Commun.} {\bf 143} (2002) 305--331,
\href{http://www.arXiv.org/abs/hep-ph/0104145}{{\tt hep-ph/0104145}}.

\bibitem{Allanach:2009bv}
B.~Allanach and M.~Bernhardt, ``{Including R-parity violation in the numerical
  computation of the spectrum of the minimal supersymmetric standard model:
  SOFTSUSY},'' {\em Comput.Phys.Commun.} {\bf 181} (2010) 232--245,
\href{http://www.arXiv.org/abs/0903.1805}{{\tt 0903.1805}}.

\bibitem{Fonseca:2011sy}
R.~M. Fonseca, ``{Calculating the renormalisation group equations of a SUSY
  model with Susyno},'' {\em Comput.Phys.Commun.} {\bf 183} (2012) 2298--2306,
\href{http://www.arXiv.org/abs/1106.5016}{{\tt 1106.5016}}.

\bibitem{Staub:2008uz}
F.~Staub, ``{SARAH},''
\href{http://www.arXiv.org/abs/0806.0538}{{\tt 0806.0538}}.

\bibitem{Staub:2009bi}
F.~Staub, ``{From Superpotential to Model Files for FeynArts and
  CalcHep/CompHep},'' {\em Comput. Phys. Commun.} {\bf 181} (2010) 1077--1086,
\href{http://www.arXiv.org/abs/0909.2863}{{\tt 0909.2863}}.

\bibitem{Staub:2010jh}
F.~Staub, ``{Automatic Calculation of supersymmetric Renormalization Group
  Equations and Self Energies},'' {\em Comput.Phys.Commun.} {\bf 182} (2011)
  808--833,
\href{http://www.arXiv.org/abs/1002.0840}{{\tt 1002.0840}}.

\bibitem{Staub:2012pb}
F.~Staub, ``{SARAH 3.2: Dirac Gauginos, UFO output, and more},'' {\em Computer
  Physics Communications} {\bf 184} (2013) pp. 1792--1809,
\href{http://www.arXiv.org/abs/1207.0906}{{\tt 1207.0906}}.

\bibitem{Machacek:1983tz}
M.~E. Machacek and M.~T. Vaughn, ``{Two Loop Renormalization Group Equations in
  a General Quantum Field Theory. 1. Wave Function Renormalization},'' {\em
  Nucl. Phys.} {\bf B222} (1983)
83.

\bibitem{Machacek:1983fi}
M.~E. Machacek and M.~T. Vaughn, ``{Two Loop Renormalization Group Equations in
  a General Quantum Field Theory. 2. Yukawa Couplings},'' {\em Nucl. Phys.}
  {\bf B236} (1984)
221.

\bibitem{Machacek:1984zw}
M.~E. Machacek and M.~T. Vaughn, ``{Two Loop Renormalization Group Equations in
  a General Quantum Field Theory. 3. Scalar Quartic Couplings},'' {\em Nucl.
  Phys.} {\bf B249} (1985)
70.

\bibitem{Jack:1982hf}
I.~Jack and H.~Osborn, ``{Two Loop Background Field Calculations for Arbitrary
  Background Fields},'' {\em Nucl.Phys.} {\bf B207} (1982)
474.

\bibitem{Jack:1982sr}
I.~Jack and H.~Osborn, ``{General Two Loop Beta Functions for Gauge Theories
  With Arbitrary Scalar Fields},'' {\em J.Phys.} {\bf A16} (1983)
1101.

\bibitem{Jack:1984vj}
I.~Jack and H.~Osborn, ``{General Background Field Calculations With Fermion
  Fields},'' {\em Nucl.Phys.} {\bf B249} (1985)
472.

\bibitem{Luo:2002ey}
M.~X. Luo and Y.~Xiao, ``{Two-loop renormalization group equations in the
  standard model},'' {\em Phys. Rev. Lett.} {\bf 90} (2003) 011601,
\href{http://www.arXiv.org/abs/hep-ph/0207271}{{\tt hep-ph/0207271}}.

\bibitem{Wingerter:2011dk}
A.~Wingerter, ``{Implications of the Stability and Triviality Bounds on the
  Standard Model with Three and Four Chiral Generations},'' {\em Phys.Rev.}
  {\bf D84} (2011) 095012,
\href{http://www.arXiv.org/abs/1109.5140}{{\tt 1109.5140}}.

\bibitem{new_sarah}
F.~Staub, ``{SARAH 4: A tool for -- not only SUSY -- model builders},'' {\em in
  prep.} (2013).

\bibitem{Cheung:2012nb}
C.~Cheung, M.~Papucci, and K.~M. Zurek, ``{Higgs and Dark Matter Hints of an
  Oasis in the Desert},'' {\em JHEP} {\bf 1207} (2012) 105,
\href{http://www.arXiv.org/abs/1203.5106}{{\tt 1203.5106}}.

\bibitem{Giudice:2004tc}
G.~Giudice and A.~Romanino, ``{Split supersymmetry},'' {\em Nucl.Phys.} {\bf
  B699} (2004) 65--89,
\href{http://www.arXiv.org/abs/hep-ph/0406088}{{\tt hep-ph/0406088}}.

\bibitem{Pyrate13}
F.~Lyonnet, I.~Schienbein, F.~Staub, and A.~Wingerter, ``The {PyR@TE} web
  page,'' 2013.
\newblock \url{http://pyrate.hepforge.org}.

\bibitem{Python}
G.~van Rossum and J.~de~Boer, ``Interactively testing remote servers using the
  python programming language,'' {\em CWI Quarterly} {\bf 4} (1991) 283--303.

\bibitem{dubois.hinsen.hugunin-1996-cp}
P.~F. Dubois, K.~Hinsen, and J.~Hugunin, ``Numerical python,'' {\em Computers
  in Physics} {\bf 10} (May/June, 1996).

\bibitem{scipy:2013}
E.~Jones, T.~Oliphant, P.~Peterson, {\em et al.}, ``{SciPy}: Open source
  scientific tools for {Python}.'' \url{http://www.scipy.org}.

\bibitem{sympy:2013}
{SymPy Development Team}, ``Sympy: Python library for symbolic mathematics.''
  \url{http://www.sympy.org}.

\bibitem{PER-GRA:2007}
F.~P\'erez and B.~E. Granger, ``{IP}ython: a {S}ystem for {I}nteractive
  {S}cientific {C}omputing,'' {\em {C}omput. {S}ci. {E}ng.} {\bf 9} (May, 2007)
  21--29.

\bibitem{PyYAML:2013}
K.~Simonov, ``Pyyaml.'' \url{http://pyyaml.org/wiki/PyYAML}.

\bibitem{yaml:2013}
C.~Evans, ``{YAML Ain't Markup Language}.'' \url{http://www.yaml.org}, 2001.

\bibitem{Holdom:1985ag}
B.~Holdom, ``{Two U(1)'s and Epsilon Charge Shifts},'' {\em Phys. Lett.} {\bf
  B166} (1986)
196.

\bibitem{Fonseca:2013bua}
R.~M. Fonseca, M.~Malinsky, and F.~Staub, ``{Renormalization group equations
  and matching in a general quantum field theory with kinetic mixing},''
\href{http://www.arXiv.org/abs/1308.1674}{{\tt 1308.1674}}.

\bibitem{Branco:2011iw}
G.~Branco, P.~Ferreira, L.~Lavoura, M.~Rebelo, M.~Sher, {\em et al.}, ``{Theory
  and phenomenology of two-Higgs-doublet models},'' {\em Phys.Rept.} {\bf 516}
  (2012) 1--102,
\href{http://www.arXiv.org/abs/1106.0034}{{\tt 1106.0034}}.

\bibitem{Basso:2010jm}
L.~Basso, S.~Moretti, and G.~M. Pruna, ``{A Renormalisation Group Equation
  Study of the Scalar Sector of the Minimal B-L Extension of the Standard
  Model},'' {\em Phys.Rev.} {\bf D82} (2010) 055018,
\href{http://www.arXiv.org/abs/1004.3039}{{\tt 1004.3039}}.

\bibitem{Arina:2012fb}
C.~Arina, J.-O. Gong, and N.~Sahu, ``{Unifying darko-lepto-genesis with scalar
  triplet inflation},'' {\em Nucl.Phys.} {\bf B865} (2012) 430--460,
\href{http://www.arXiv.org/abs/1206.0009}{{\tt 1206.0009}}.

\bibitem{Pirogov:1998tj}
Y.~F. Pirogov and O.~V. Zenin, ``{Two-loop renormalization group restrictions
  on the standard model and the fourth chiral family},'' {\em Eur. Phys. J.}
  {\bf C10} (1999) 629--638,
\href{http://www.arXiv.org/abs/hep-ph/9808396}{{\tt hep-ph/9808396}}.

\bibitem{Ishiwata:2011hr}
K.~Ishiwata and M.~B. Wise, ``{Higgs Properties and Fourth Generation
  Leptons},'' {\em Phys.Rev.} {\bf D84} (2011) 055025,
\href{http://www.arXiv.org/abs/1107.1490}{{\tt 1107.1490}}.

\end{thebibliography}\endgroup
\bibliographystyle{styles/utphys}

\end{document}